\documentclass[a4paper,fleqn,usenatbib]{mnras}

\usepackage[T1]{fontenc}
\usepackage{ae,aecompl}
\usepackage{bm}

\usepackage{amsmath}	
\usepackage{amssymb}	
\usepackage{graphicx}
\usepackage{wrapfig}

\newcommand{\Msun}{{\rm M_\odot}}
\newcommand{\Zsun}{{\rm Z_\odot}}
\newcommand{\fdense}{{f_{\rm dense}}}
\newcommand{\Dtot}{{\mathcal{D}_\mathrm{tot}}}
\newcommand{\DS}{{\mathcal{D}_\mathrm{S}}}
\newcommand{\DL}{{\mathcal{D}_\mathrm{L}}}
\newcommand{\DSL}{{\DS/\DL}}
\newcommand{\eSN}{{\varepsilon_{\rm SN}}}

\title[Dust formation and destruction in galaxies]{Galaxy Simulation with Dust Formation and Destruction}

\author[S. Aoyama et al.]{Shohei Aoyama,$^{1,2}$\thanks{E-mail: aoyama@vega.ess.sci.osaka-u.ac.jp (SA)}
Kuan-Chou Hou,$^{3,4}$
Ikkoh Shimizu,$^{1,2}$
Hiroyuki Hirashita,$^{3}$\newauthor
Keita Todoroki,$^{5,6}$ 
Jun-Hwan Choi,$^{7}$ and 
Kentaro Nagamine$^{1,6}$\\
$^{1}$Theoretical Astrophysics, Department of Earth \& Space Science, Osaka University, 1-1 Machikaneyama, Toyonaka, Osaka 560-0043, Japan\\
$^{2}$College of General Education, Osaka Sangyo University, 3-1-1, Nakagaito, Daito, Osaka, 574-8530, Japan\\
$^{3}$Institute of Astronomy, and Astrophysics, Academia Sinica, PO Box 23-141, Taipei 10617, Taiwan\\
$^{4}$Department of Physics \& Institute of Astrophysics, National Taiwan University, Taipei 10617, Taiwan \\
$^{5}$Department of Physics \& Astronomy, University of Kansas,1082 Malott,1251 Wescoe Hall Dr., Lawrence, KS 66045-758, USA\\
$^{6}$Department of Physics \& Astronomy, University of Nevada, Las Vegas, 4505 S. Maryland Pkwy, Las Vegas, NV 89154-4002, USA \\
$^{7}$Department of Astronomy, University of Texas Austin, TX 78712-1205, USA
}

\date{Accepted 2016 November 22. Received 2016 November 18; in original form 2016 September 22}

\pubyear{2017}

\begin{document}
\label{firstpage}
\pagerange{\pageref{firstpage}--\pageref{lastpage}}
\maketitle

\begin{abstract}
We perform smoothed particle hydrodynamics (SPH) simulations of an isolated galaxy 
with a new treatment for dust formation and destruction. 
To this aim, we treat dust and metal production self-consistently 
with star formation and supernova (SN) feedback. For dust, we consider a simplified
model of grain size distribution by representing the entire range of grain sizes
with large and small grains.
We include dust production in stellar ejecta, dust destruction
by SN shocks, grain growth by accretion and coagulation, 
and grain disruption by shattering.
We find that the assumption of fixed dust-to-metal mass ratio becomes 
no longer valid when the galaxy is older than 0.2\,Gyr, 
at which point the grain growth by accretion starts to contribute to the nonlinear rise of dust-to-gas ratio.
As expected in our previous one-zone model, shattering triggers grain
growth by accretion since it increases the total surface area of grains.
Coagulation becomes significant when the galaxy age is greater than $\sim$\,1\,Gyr: 
at this epoch the abundance of small grains becomes high enough to raise the coagulation
rate of small grains. 
We further compare the radial profiles of dust-to-gas ratio $(\mathcal{D})$
and dust-to-metal ratio $(\mathcal{D}/Z)$
(i.e., depletion) at various ages with observational data.
We find that our simulations broadly reproduce
the radial gradients of dust-to-gas ratio and depletion.
In the early epoch ($\lesssim 0.3$\,Gyr), 
the radial gradient of $\mathcal{D}$ follows 
the metallicity gradient with $\mathcal{D}/Z$ determined by 
the dust condensation efficiency 
in stellar ejecta, while the $\mathcal{D}$ gradient is steeper than
the $Z$ gradient at the later epochs because of grain growth by accretion.
The framework developed in this paper is applicable to 
any SPH-based galaxy evolution simulations including cosmological ones.
\end{abstract}




\section{Introduction}
The importance of cosmic dust in astrophysical processes has been recognized in recent decades.
Dust acts as an efficient catalyst 
of molecular hydrogen (H$_2$) formation 
in the interstellar medium \citep[ISM; e.g.][]{1963ApJ...138..393G,2004ApJ...604..222C,2009A&A...496..365C}.
In addition, the typical mass of the final fragments in star-forming clouds 
is also regulated by dust cooling \citep[][]{1998MNRAS.299..554W,2005MNRAS.359..211L,2005ApJ...626..627O,2006MNRAS.369.1437S}.
In protoplanetary discs, dust growth eventually leads to planet formation
\citep[e.g.,][]{2009ApJ...707.1247O,2014A&A...568A..42K}.
Dust grains also play an important role in radiative processes in the ISM by
absorbing stellar light and reemitting it 
in the far-infrared, and change the spectral energy distributions of galaxies 
\citep[][]{2000ApJ...533..682C,2002A&A...383..801B,2012ApJ...755..144T}.

For the dust properties in galaxies, 
the size distribution of dust grains is of fundamental importance
\citep[e.g.,][]{1977ApJ...217..425M,2013ApJ...770...27N}. 
In particular, the extinction curve 
(i.e., the wavelength dependence of absorption and scattering cross-section)
is shaped by the grain size distribution, given the grain materials
\citep[][]{1983Natur.306..625B}.
Precise estimates of star formation rate (SFR) 
in galaxies also require correction for dust extinction
\citep[e.g.,][]{1999ApJ...519....1S,2010A&A...514A...4T,2012ARA&A..50..531K}. 
In addition, the total dust surface area which depends on the grain size distribution,
governs the formation rate of molecular hydrogen 
\citep[e.g.,][]{1976ApJ...207..131B,2011ApJ...735...44Y}.

Dust interacts with gas, metals and dust itself in the ISM.
It is not only destroyed by supernova (SN) shocks,
but also disrupted or shattered by grain--grain collisions
in the diffuse ISM \citep[][]{2004ApJ...616..895Y}.
In dense environments such as molecular clouds, 
it grows by accretion and coagulation \citep[][]{2014MNRAS.437.1636H,2014MNRAS.445..301V}.
All these processes play important roles in the evolution of 
dust abundance and grain size distribution.

\cite{2013MNRAS.432..637A} have established a full framework 
for treating the evolution of grain size distribution 
consistently with the chemical enrichment in a galaxy.
Their work revealed that the collisional effects of dust grains
such as coagulation, shattering, and accretion 
are necessary for a comprehensive understanding
of the observed dust-to-gas mass ratio and extinction curves in nearby galaxies.
However, \cite{2013MNRAS.432..637A} treated a galaxy as a single zone
without taking into account the spatial distribution of gas 
with different density structures. 
Since the efficiencies of various dust processing mechanisms
depend on the density and temperature of the ISM, 
hydrodynamical evolution of the ISM should also be considered
simultaneously with dust evolution.

Hydrodynamical simulations have indeed been 
a powerful tool to clarify galaxy formation and evolution.
Many cosmological hydrodynamic simulations have reproduced
 and predicted the observed galaxy mass and luminosity functions 
\citep[e.g.,][]{2001ApJ...558..497N,2004MNRAS.350..385N,2012MNRAS.419.1280C,2012MNRAS.427..403J,2013ApJ...766...94J,2014MNRAS.440..731S,2014ApJ...780..145T,2014MNRAS.444.1518V,2015arXiv150900800S,2015MNRAS.446..521S}.
In order to compute luminosity functions from simulated galaxies and 
compare them with observed data,  
a precise estimate of dust extinction effect is required. 
For example, \cite{2015MNRAS.451..418Y} calculated the galaxy UV luminosity function 
at high redshifts $(6 \le z \le 12)$ using 
cosmological zoom-in hydrodynamical  simulations with constrained initial conditions. 
In their work, chemistry and cooling of hydrogen, helium and metals were 
computed as in \cite{2009MNRAS.393.1595C}, 
however, the dust-to-metal ratio was fixed.
\cite{2015MNRAS.449.1625B} treated dust 
as a new particle species in addition to gas, dark matter, and star particles. 
They included not only the formation and destruction of dust,  
but also dust-dependent star formation and stellar feedback 
\citep[][]{2013MNRAS.432.2298B}.
Furthermore, in a more recent work, 
\cite{2016MNRAS.457.3775M} regarded dust as 
an additional component in gas, and performed 
cosmological zoom-in simulations. 
They revealed the importance of
dust growth by the accretion of gas-phase metals, and 
pointed out the necessity of a more realistic treatment 
of dust destruction and feedback by SNe. 
In addition, \citet[][]{2016arXiv160602714M} ran cosmological simulations
and compared the dust mass function and radial profile of dust 
with corresponding observational data. 
They found that their simulation broadly reproduced the observation in the present-day Universe,
although it tended to underestimate the dust abundance in high-redshift galaxies.
\textcolor{black}{We also note that a similar approach on dust models can be taken with a semi-analytic model of galaxy formation. For example, \citet[][]{2016arXiv160908622P} treated dust growth in dense ISM, destruction by SN shocks and chemical evolution, and estimated the dust abundance and dust mass function of galaxies at various redshifts using a semi-analytic model. 
}

In all of these simulations mentioned above,
dust processing by grain-grain collisions 
such as coagulation and shattering, 
both of which are important for the grain size distribution,
has not been included yet \citep[but see][who implemented part of these processes by postprocessing a hydrodynamic simulation of an isolated galaxy with a 
particular focus on temperature-dependent sticking coefficient]{2016arXiv160804781Z}. 
Implementation of dust size distributions in smoothed particle hydrodynamics
(SPH) simulations has not been successful,
mainly because of the high computational cost.
Calculating the grain size distribution in a fully self-consistent manner 
is computationally expensive even in one-zone calculation as shown by \citet[][]{2013MNRAS.432..637A}. 
However, because of the aforementioned effects of grain size distribution,
implementation of grain size distribution in hydrodynamic simulations is essential
in understanding the role of dust in galaxy evolution.

In this paper, we perform $N$-body/SPH simulations 
of isolated galaxies with a model of dust formation and destruction. 
For the purpose of treating the evolution of grain size distribution 
within a reasonable computational time,
we adopt the two-size approximation formulated by \cite{2015MNRAS.447.2937H}, 
in which the entire grain size range is represented by two sizes 
divided at around $a\simeq 0.03~\mu$m, where $a$ is the grain radius.
\cite{2015MNRAS.447.2937H} has confirmed that the two-size approximation 
gives the same evolutionary behaviour of 
grain sizes and extinction curves as calculated by the full treatment of 
\cite{2013MNRAS.432..637A,2014MNRAS.440..134A}.
Thus, implementing the two-size model, 
which is computationally light, in hydrodynamic simulations 
provides a feasible way to compute the grain size evolution in the ISM.
Consequently, not only can we compute the spatial variations in 
dust properties, but also examine the grain size distribution as a function 
of time and metallicity. 
This is a significant advantage over the simple one-zone calculations,
which generally need to introduce some strong assumptions such as
instantaneous mixing and homogeneity.

Although our ultimate goal is to understand dust evolution in
cosmological structure formation,
the target of this paper is an isolated galaxy for the purpose of
the first implementation of  dust evolution. 
Using an isolated galaxy enables us to compare our results with
previous one-zone calculations and to test our implementation.
Since the spatial resolution is higher than typical cosmological simulations,
we will be able to predict spatially resolved properties of dust
evolution in detail.

This paper is organized as follows. 
In Section 2, we introduce our dust evolution model and 
calculation method. We present the simulation results in Section 3. 
We discuss the parameter dependence in Section 4
and compare the simulation results with observational data in Section 5. 
We conclude in Section 6.
Throughout this paper, 
we adopt $Z_{\odot}=0.02$ for solar metallicity 
following \cite{2015MNRAS.447.2937H}.

\begin{table}
\caption{Initial physical parameters of our isolated galaxy. 
In this paper, we adopt the low-resolution model of the {\sf AGORA} project \citep{2014ApJS..210...14K}.
The disc and bulge components are pre-existing stellar components treated by collisionless star  particles dynamically,  but those particles are not destroyed nor created during the simulation. 
$^\dagger$The gravitational softening length is taken to be 80 pc, and we allow the baryons to collapse to 10\% of this value.  However, in practice, we find that the variable gas smoothing length reached a minimum value of only $\sim$\,22 pc with our models of gas cooling, star formation, and feedback. 
}
\begin{center}
  \begin{tabular}{lcc}\hline
Parameter & Value & \\ \hline
Gas mass  & $8.59 \times 10^{9} \Msun$ \\ 
Dark matter mass & $1.25 \times 10^{12} \Msun$ \\
Disc mass & $4.30 \times 10^{9} \Msun$ \\
Bulge mass & $3.44 \times 10^{10} \Msun$ \\  \hline
Total mass & $1.3 \times 10^{12} \Msun$ \\ \hline
Number of gas particle & $1.00\times 10^{5}$ \\
Number of dark matter & $1.00\times 10^{5}$ \\
Number of disc particle & $1.00\times 10^{5}$ \\
Number of bulge particle & $1.25\times 10^{4}$ \\ \hline 
Gas particle mass & $8.59 \times 10^{4} \Msun$ \\
Dark matter particle mass & $1.25 \times 10^{7} \Msun$ \\
Disc particle mass & $3.44 \times 10^{5} \Msun$ \\
Bulge particle mass & $3.44 \times 10^{5} \Msun$ \\ \hline 
Grav. softening length     &  $80$ pc $^\dagger$\\ 
\hline
\end{tabular}
\label{table:params}
\end{center}
\end{table}


\section{Model}

\subsection{Numerical Simulations}

In this section, we overview our simulation setup first before 
we describe the details of our dust implementation. 
We use the modified version of {\tt GADGET-3} 
$N$-body/SPH code \citep[originally explained by][]{2005MNRAS.364.1105S}, and 
our simulation includes dark matter, gas, and star particles. 
The dynamics of collisionless particles are computed with a 
tree-particle-mesh method, and the hydrodynamics is solved with the
entropy-conserving, density-independent formulation 
\citep{2002MNRAS.333..649S,2013ApJ...768...44S,2013MNRAS.428.2840H},  
using the quintic spline kernel of \citet{1996PASA...13...97M}.
The star particles are stochastically created 
from gas particles as described in \cite{2003MNRAS.339..289S}, 
consistently with the SFR equation that we describe below. 
Our code uses the Grackle\footnote{https://grackle.readthedocs.org/} 
chemistry and cooling library 
\citep{2014ApJS..211...19B,2014ApJS..210...14K},
which solves the non-equilibrium primordial chemistry network 
for atomic H, D, He, H$_2$ and HD. 
Using the Grackle's non-equilibrium chemistry with H$_2$ 
allows us to compute the gas properties to lower temperatures and higher densities, which is necessary for a proper treatment of dust physics as we will see below.  

The initial condition of our isolated galaxy is taken from the 
low-resolution model of {\sf AGORA} simulations \citep[][]{2014ApJS..210...14K}, 
and various parameter values of this disc galaxy are summarized in 
Table~\ref{table:params}. 
The highest density that we can resolve is also determined by the minimum gravitational softening length of $\epsilon_{\rm grav}=80$\,pc, 
and we allow the minimum gas smoothing to reach 10\% of $\epsilon_{\rm grav}$.
Following \citet{2011MNRAS.417..950H} and Kim et al. (2016), 	
we adopt the Jeans pressure floor, 
\begin{equation}
P_{\rm Jeans} = 1.2 \gamma^{-1} N_{\rm Jeans}^{2/3} G \rho_{\rm gas}^2 h_{\rm sml}^2,
\end{equation}
where $\gamma=5/3$, $N_{\rm Jeans}=8.75$, and $h_{\rm sml}$ is the 
smoothing length of gas. 
This is to avoid artificial numerical fragmentation when the Jeans mass 
at low temperatures is not resolved. 

The gas particles in our simulations carry physical information
such as density, internal energy, entropy, ionization fraction, 
metallicity, and SFR. 
The gas pressure can be computed from entropy and density, 
and the temperature from internal energy and ionization fraction. 
Therefore we can use these dynamically computed physical 
quantities to evaluate the dust processes at each point in space and time. 
With our new treatment for dust, the gas particles now also carry 
the dust mass as an additional physical parameter. 

\begin{figure}
\includegraphics[width=8.3cm]{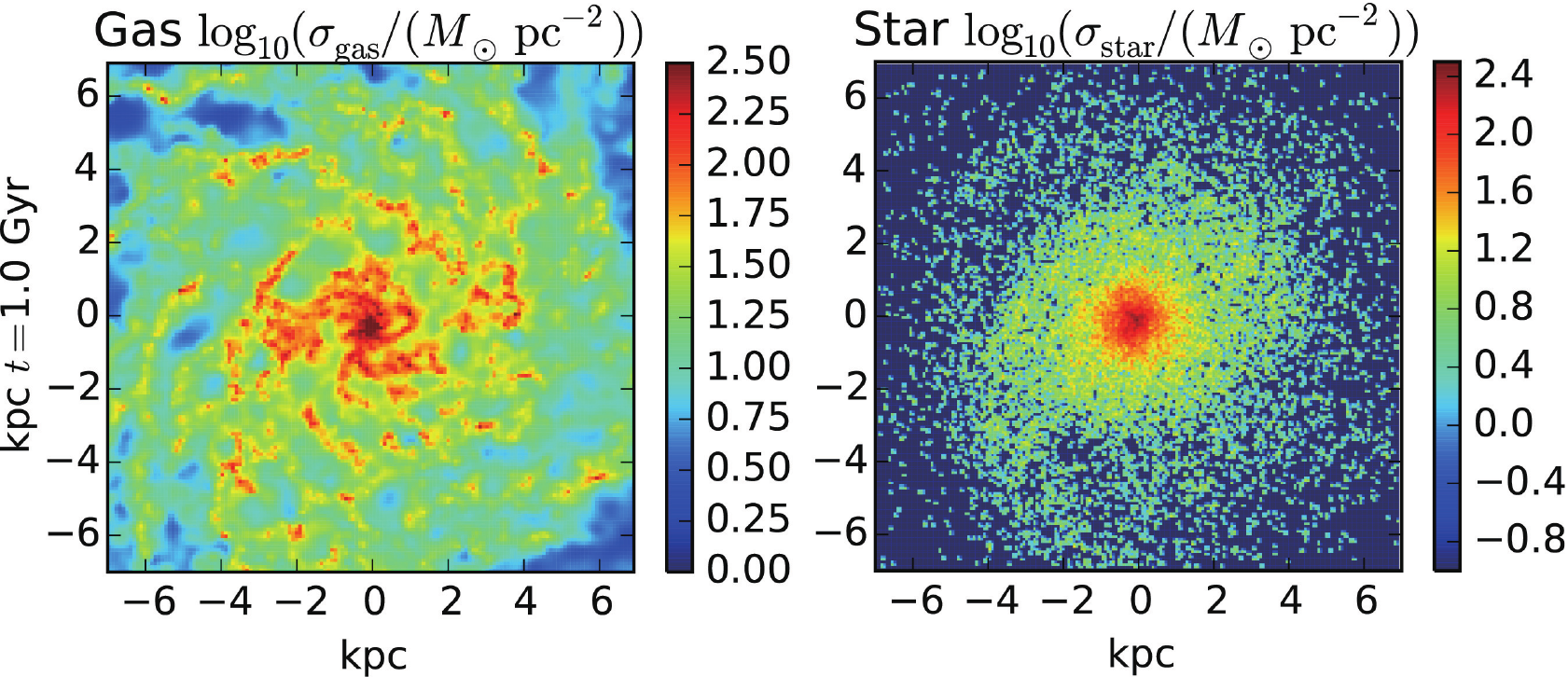}
\caption{Face-on view of the distribution of 
gas surface density (left-hand panel) and stellar surface density (right) of 
our simulation at $t=1$ Gyr. 
The colour bar indicates the mass surface density 
in logarithmic units of [$\Msun$\,pc$^{-2}$].
The image was produced with a pixel size of $\sim$\,93\,pc. 
}
\label{fig:gas_star_surface}
\end{figure}

As an example, Fig.~\ref{fig:gas_star_surface} shows the face-on view 
of surface densities of gas and stars of our simulated galaxy 
at $t=1$ Gyr.   Spiral arms are prominent and well developed after 1\,Gyr
since the beginning of the simulation, and stellar disc is clearly visible 
in the right-hand panel.  
We also smooth the spatial distribution of metal and dust within the 
smoothing kernel, in a similar manner to the computation of gas density
while the simulation is running.

\begin{figure}
\includegraphics[width=8.5cm]{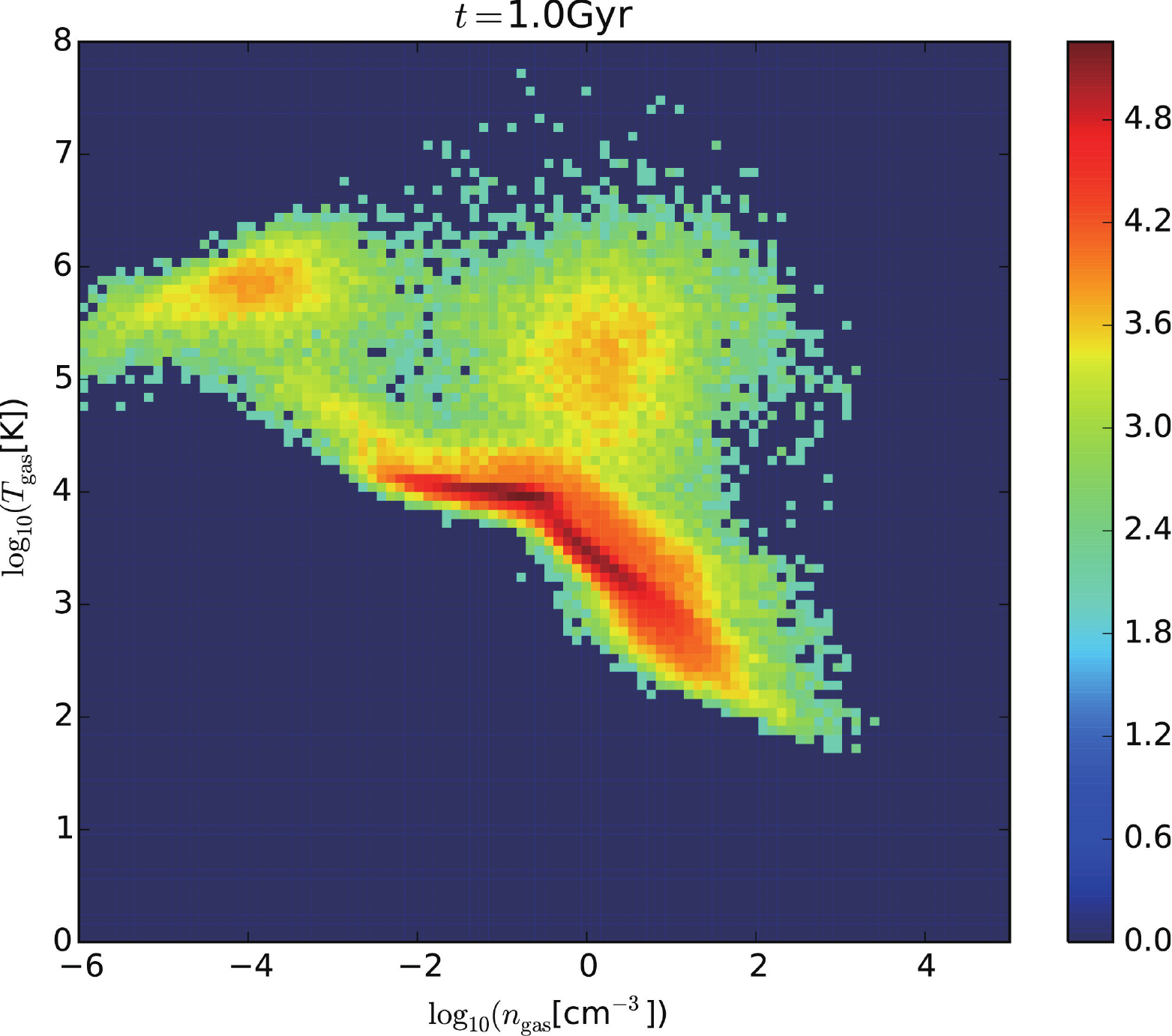}
\caption{Distribution of gas particles on the temperature--density  
($T_{\rm gas}$--$n_{\rm gas}$) phase diagram in our simulation at $t=1$ Gyr.
The colour indicates the logarithmic surface density of gas particles on this diagram
as indicated by the colour bar on the right.
}
\label{fig:phase}
\end{figure}

Fig.~\ref{fig:phase} shows the overall distribution of gas 
in our simulated galaxy on the temperature--density plane. 
With our current resolution and Grackle cooling, 
we achieve gas number densities slightly higher than 10$^4$\,cm$^{-3}$
and temperatures as low as $T_{\rm gas} \sim 30$\,K. 
There is also a distribution of tenuous gas heated by the stellar feedback effects 
(see Section~\ref{sec:stellarfb}) at $T_{\rm gas} > 10^{5}$\,K, 
which would be absent if we turn off the feedback processes. 

\subsubsection{Star Formation}

We follow the star formation prescription of the {\sf AGORA} project 
\citep{2014ApJS..210...14K}. 
In this model, the local SFR density $d\rho_{\ast }\slash dt$ is estimated as 
\begin{eqnarray} 
\frac{d \rho_{\ast}}{dt} = \varepsilon_{\rm SF}\frac{\rho_{\rm gas}}{t_{\rm ff}},
\end{eqnarray}
where $\varepsilon_{\rm SF}=0.01$ is the star formation efficiency, 
$\rho_{\rm gas}$ is the gas density and 
$t_{\rm ff}=\sqrt{3\pi / (32G\rho_{\rm gas})}$
is the local free-fall time. 
The star particles are stochastically created 
from gas particles as described in \cite{2003MNRAS.339..289S}, 
consistently with the above SFR. 
But the star formation is allowed only for gas particles 
with $n_{\rm gas} \ge 10~{\rm cm}^{-3}$. 
Each star particle is regarded as a simple stellar population 
with a \citet{2003PASP..115..763C} initial stellar mass function (IMF), 
and it carries stellar mass, metallicity, and formation time 
as its physical attributes. 

Our simulation results show that the gas is converted into stars at a steady rate of 
$\sim 1\,\Msun$ yr$^{-1}$, although we do not show the figure here as it is a simple result. 
Total stellar mass increases monotonically
from zero at $t=0$ Gyr to $3.4\times 10^{9} \Msun$ at $t=10$ Gyr. 
Disc and bulge mass given in Table~\ref{table:params} do not change
as they are assumed to be made of old stellar population and treated by 
collisionless stellar particles.  The overall gas mass of the galaxy decreases
monotonically from $8.6\times 10^{9}\,\Msun$ at $t=0$ Gyr to $5.2\times 10^{9}\,\Msun$ 
at $t=10$ Gyr.


\subsubsection{Early Stellar Feedback}
\label{sec:stellarfb}

Several studies have pointed out the impact of feedback 
from stellar winds and radiation of massive stars before they 
explode as SNe \citep[][]{2012MNRAS.427..311W,2013ApJ...770...25A,2013MNRAS.428..129S}, 
which is often called the `early stellar feedback.'  
For example, \citet[][]{2013ApJ...770...25A} 
considered the momentum of ionizing radiation on to the ambient gas, 
and found that it is effective in sweeping up the gas 
in the vicinity of young star clusters. 
\citet{2013MNRAS.428..129S} also implemented a 
phenomenological model, which 
assumes that the energy from massive stars are converted and 
deposited among the neighbouring gas particles 
in the form of pure thermal energy, while turning the 
radiative cooling off for 30\,Myr to avoid the overcooling problem. 

While this work is more focused on the physics of dust 
formation and evolution, we have implemented the effect of 
stellar feedback model, roughly following the work by \citet{2013MNRAS.428..129S} and 
\cite{2014MsT..........1T}. 
In our model, the thermal energy from early stellar feedback is injected 
at a constant rate for $t_{\mathrm{explode}} = 4$\,Myr, 
which is an approximate lifetime of a massive star 
\citep[e.g.,][]{2010ApJ...710L.142F}. 
The fractional thermal energy deposited from a star particle in a single time-step is  
\begin{eqnarray}
\Delta{E_{\rm th}} = \epsilon_{\rm esf} \ E_{\rm rad} \ \frac{t - t_{\rm deposit}}{t_{\rm explode}}, 
\end{eqnarray}
where $t$ is the current simulation time,
$t_{\mathrm{deposit}}$ is the time at which the star deposited a fraction of 
the thermal energy in the previous time-step,
$\epsilon_{\rm esf}$ is the early stellar feedback efficiency, and
$E_{\rm rad}$ is the radiation energy that is emitted by massive stars. 
We adopt $E_{\rm rad}=2\times 10^{50}$ erg per 1\,$\Msun$ of stars \citep[][]{2013MNRAS.428..129S}.

The energy $\Delta E_{\mathrm{th}}$ is then distributed among all the neighbour gas particles using the mass-weighted SPH kernel. 
The physical meaning of early stellar feedback efficiency, $\epsilon_{\rm esf}$, is the fraction of UV photons that contributes to the early stellar feedback. 
\citet{2013MNRAS.428..129S} find that the model with 
$\epsilon_{\rm esf}= 0.1$ reproduces a good agreement with the stellar-to-halo mass relation obtained from halo occupation distribution models, but here we take $\epsilon_{\rm esf} = 0.3$ to see the maximum impact of early stellar feedback. 
However, having examined our results on dust properties 
in the runs with and without the early stellar feedback, 
we actually find that its impact on dust results is not so strong.


\subsubsection{SN model}
\label{SNeDestruction4.3.2}

Dust destruction by SNe is also treated by a subgrid model, as our simulations
do not have the resolution to solve the gas dynamics at sub-parsec scales. 
In our subgrid model, we first compute a `shock radius' ($R_\mathrm{sh}$) for newly formed star particles, 
inside which the gas particles are affected by the SN shock and the dust is 
destroyed. 
We adopt the shock radius from \citet[][]{1974ApJ...188..501C} and \citet[][]{1977ApJ...218..148M}, 
who performed spherically symmetric hydrodynamical simulation 
of SN remnant in a uniform medium.
In their simulation, they considered not only the cooling via infrared, UV and X-ray radiation, 
but also the effects of magnetic field on the remnant. 
As a consequence, they obtained a shock radius as
\begin{eqnarray} 
R_{\rm sh}&=&55.0~[{\rm pc}] \left(\dfrac{E_{\rm SN}}{10^{51}~{\rm erg}} \right)^{0.32} \notag \\
& &\times \left(\dfrac{n_{\rm gas}}{1~{\rm cm}^{-3}} \right)^{-0.16}
\left(\dfrac{P_{\rm 0}}{10^{-4}k_{\rm B}~{\rm cm}^{-3}\,{\rm K}} \right)^{-0.20}~,
\end{eqnarray}
where $P_{\rm 0}$ is the ambient pressure that is obtained by calculating 
the kernel-weighted mean pressure of neighbouring gas particles 
located within the smoothing length of the young star particle. 
The value of $E_{\rm SN}$, which is the total collective energy from 
all SNe in the star particle of concern, will be specified in equation \eqref{eq:Esn}.


\subsubsection{Metal production and SN feedback}

Dust is produced from metals, and metals are ejected from SN explosions; 
therefore we need to treat dust and metal production 
in a consistent manner with star formation and SN feedback. 
In this subsection, we first describe our treatment of SN feedback 
and metal enrichment, which are based on the 
{\sf AGORA} project \citep{2014ApJS..210...14K} and \citet{2014MsT..........1T}.

We assume that stars with masses between 8 and $40 \Msun$ 
explode as Type II SNe  after a time-delay of 
$t_{\mathrm{explode}} = 4$\,Myr from the star formation, 
and deposit a net energy of $10^{51}$\,erg into the ambient medium.  
For the assumed Chabrier IMF, 
the number of Type II SNe per unit stellar mass is 0.011\,$\Msun^{-1}$. 
We include the effects of gas recycling and metal production from Type II SNe.  
We return the mass of stars in between $8 \Msun < M < 40 \Msun$ into ISM
after subtracting the remnant mass of 1.4 $\Msun$ per star.

The total amount of metals produced and ejected from Type II SNe per stellar mass, 
$M_{\rm Z}$, is calculated as
\begin{eqnarray}
M_{\rm Z} = 2.09M_{\rm O} + 1.06M_{\rm Fe}, 
\label{eq:metalproduce}
\end{eqnarray}
with the corresponding fractional masses of oxygen $M_{\rm O}$ and iron $M_{\rm Fe}$ 
being 0.0133 $\Msun$ and 0.011 $\Msun$, respectively, per 
one solar mass of stars formed. 
These values come from the tabulated results of  
\citet{2007PhR...442..269W} and for the Chabrier IMF 
\citep[see also section 3.5 of][]{2014ApJS..210...14K}. 
Inserting these values into equation.~(\ref{eq:metalproduce}) gives
$M_{\rm Z} = 0.02896\,\Msun$ per one solar mass of stars formed. 
In other words, in the actual simulation, 
the amount of metal mass injected into the $i$-th gas particle
is computed as 
\begin{equation} 
\Delta A_{i} = \frac{m_{i} W(|\bm{r}_{i} - \bm{r}_{\rm s}|, h_{\rm s})}{\sum\limits_{j=1}^{N} m_{j} W(|\bm{r}_{j} - \bm{r}_{\rm s} |, h_{\rm s} )} \ A, 
\label{eq:metalproduce2}
\end{equation}
where $A$ is taken as $\mathcal{Y}^{\prime }_{\rm Z} \Delta \tilde{m}_{\ast} $ in this case, 
$\mathcal{Y}^{\prime}_{\rm Z} = 0.02896$ is the effective yield, 
$\Delta \tilde{m}_{\ast }$ is the mass of the star particle 
that provides the metal and dust to the $i$th gas particle with mass $m_i$,
$W$ is a kernel function, $\bm{r}_{i}$ is the positional vector of 
$i$th gas particle measured from the star particle of concern at $\bm{r}_{\rm s}$, 
and ${h}_{\rm s}$ is the smoothing length recomputed for the star particle
using neighbouring gas particles, respectively.

The produced metals are distributed by the SN shock wave
to the neighbouring gas particles within $R_{\rm sh}$. 
If there is only one neighbouring gas particle within $R_{\rm sh}$ 
(which could happen in a low resolution simulation), 
then all the metals are given to the corresponding gas particle. 

When a star particle with mass $m_{\ast}$ explodes,
the following energy $E_{\rm SN}$ is released \citep[][]{2014ApJS..210...14K}:
\begin{eqnarray} 
E_{\rm SN} = 0.011\times 10^{51} \left( m_{\ast} \slash \Msun \right)~{\rm erg}\, .
\label{eq:Esn}
\end{eqnarray}
As for the kinetic SN feedback,
we take 28.3\% of the total SN energy 
\citep[$E_{\rm SN,k} = 0.283 E_{\rm SN}$;][]{2012MNRAS.419..465D} 
as the available kinetic energy, 
and give the corresponding momentum to gas particles 
within $R_{\rm sh}$ in random directions. 
The remaining 71.7\% of the total SN energy is given to the gas 
as a thermal feedback similar to the metals above.

\subsubsection{Time evolution of dust abundances}

We solve dust evolution on each gas particle in our SPH simulation. 
\textcolor{black}{
At this point, our simulation code does not follow 
metal and dust evolution element by element.
It is our future plan to treat them separately coupled with SN and cooling models 
\citep[e.g.,][]{2010MNRAS.409..132W}. 
We are also preparing another paper treating carbonaceous and silicate dust separately 
to examine their impact on the shape of extinction curve (Hou et al., in preparation). 
}

In this paper, the treatment of grain size distribution is 
based on the two-size approximation by \cite{2015MNRAS.447.2937H}. 
The whole range of grain radii  
is represented by large and small grain populations 
considering that various grain processing mechanisms work
differently between these two grain populations.
The boundary of these two populations is chosen 
at $a\simeq 0.03~\mu$m \citep[][]{2015MNRAS.447.2937H}
based on the full treatment of grain size distribution 
by \cite{2013MNRAS.432..637A}.
We adopt typical radii of 
the large and small grain populations as 
$0.1~\mu$m and $5\times 10^{-3}~\mu $m, respectively. 
The abundances of the two dust populations on a gas particle
are represented by the dust-to-gas mass ratios,
$\mathcal{D}_{\rm L}$ and $\mathcal{D}_{\rm S}$ as
\begin{eqnarray} 
\mathcal{D}_{\rm L}&=&\dfrac{m_{\rm L}}{m_{\rm g}}\, ,\\
\mathcal{D}_{\rm S}&=&\dfrac{m_{\rm S}}{m_{\rm g}}\, ,
\end{eqnarray}
where $m_{\rm g}$ is the mass of the gas particle, and
$m_{\rm L}$ and $m_{\rm S}$ are the total mass of large and small grains  
in the gas particle, respectively.
Hereafter we refer to $\mathcal{D}_{\rm L}$ ($\mathcal{D}_{\rm S}$) as 
the large (small) grain abundance. 
The total dust-to-gas ratio $\mathcal{D}_{\rm tot}$ is defined as
\begin{eqnarray} 
\mathcal{D}_{\rm tot}&\equiv &\mathcal{D}_{\rm L}+\mathcal{D}_{\rm S}\, .
\end{eqnarray}
In our simulation, each gas particle has its own dust abundance 
$\mathcal{D}_{{\rm L}(i)}$ and $\mathcal{D}_{{\rm S}(i)}$, where suffix $i$ indicates
the label for the gas particle. 
The dust production and destruction processes
are calculated for each particle 
by the model described below.

Based on the two-size model,
we calculate the formation and destruction of large and small dust grains 
on each gas particle using variables and outputs in the simulation.
Considering the relevant processes for large and small grains,
the equations governing the masses of large and small grains
are written as \citep{2015MNRAS.447.2937H}
\begin{align}
\frac{\mathrm{d}m_\mathrm{S}}{\mathrm{d}t} &=
-\mathcal{D_\mathrm{S}}\psi -\frac{m_\mathrm{S}}{\tau_\mathrm{SN}}+
\frac{m_\mathrm{L}}{\tau_\mathrm{sh}}
-\frac{m_\mathrm{S}}{\tau_\mathrm{co}}
+\frac{m_\mathrm{S}}{\tau_\mathrm{acc}},\label{eq:dMdsdt}\\
\frac{\mathrm{d}m_\mathrm{L}}{\mathrm{d}t} &=
-\mathcal{D_\mathrm{L}}\psi +f_\mathrm{in}E_Z-
\frac{m_\mathrm{L}}{\tau_\mathrm{SN}}-
\frac{m_\mathrm{L}}{\tau_\mathrm{sh}}+
\frac{m_\mathrm{S}}{\tau_\mathrm{co}},\label{eq:dMdldt}
\end{align}
where $\psi$ is the SFR, $E_Z$ is
the injection rate of metals from stars, and $f_\mathrm{in}$
is the dust condensation efficiency of the metals in
the stellar ejecta. The time-scales of various processes
are also introduced: $\tau_\mathrm{SN}$ is the
time-scale of dust destruction by SN shocks, and
$\tau_\mathrm{sh}$,
$\tau_\mathrm{co}$ and $\tau_\mathrm{acc}$ are the
time-scales of shattering, coagulation and accretion,
respectively.

Based on the above equations, we formulate
the time evolution in the large and small grain abundances
in the $i$-th particle from time $t$ 
to the next time step $t + \Delta t$ as
\citet[see also equation 15 and 16 in ][]{2015MNRAS.447.2937H}
\begin{eqnarray} 
\mathcal{D}_{{\rm L}(i)}(t+\Delta t)
&=&\mathcal{D}_{{\rm L}(i)}(t) - \Delta \mathcal{D}_{({\rm SNe / L})(i)} \notag \\ 
& &- \left( \dfrac{\mathcal{D}_{{\rm L}(i)}(t)}{\tau_{\rm sh}}
- \dfrac{\mathcal{D}_{{\rm S}(i)}(t)}{\tau_{\rm co}}\right)\Delta t \notag \\
& &+ f_{\rm in }\dfrac{\Delta \tilde{m}_{\rm Z}}{m_{\rm g}}
( 1 - \delta )\, ,  \label{eq:timeL} \\ 
\mathcal{D}_{{\rm S}(i)}(t+\Delta t)
&=&\mathcal{D}_{{\rm S}(i)}(t) - \Delta \mathcal{D}_{({\rm SNe / S})(i)}  \notag \\ 
& &+ \left(\dfrac{\mathcal{D}_{{\rm L}(i)}(t)}{\tau_{\rm sh}}
- \dfrac{\mathcal{D}_{{\rm S}(i)}(t)}{\tau_{\rm co}}  
+\dfrac{\mathcal{D}_{{\rm S}(i)}(t)}{\tau_{\rm acc}}\right)\Delta t\, , \notag \\
\label{eq:timeS} 
\end{eqnarray}
where $\Delta \mathcal{D}_{{\rm (SNe / L)}(i)}$ 
and $\Delta \mathcal{D}_{{\rm (SNe / S)}(i)}$ are 
the decrease of the large and small grain abundances 
in the $i$-th particle by SN destruction, respectively.
The amount of newly produced metals, $\Delta\tilde{m}_Z$, is computed
according to equation~(\ref{eq:metalproduce2}).
 The destroyed fraction of the newly supplied dust from stars, $\delta$, 
is derived in Appendix \ref{dustDestructionRate}.
We explain how to estimate $\Delta \mathcal{D}_{\mathrm{(SNe /  L)}(i)}$, 
$\Delta \mathcal{D}_{\mathrm{(SNe / S)}(i)}$,
$\tau_{\rm sh}$, $\tau_{\rm co}$ and $\tau_{\rm acc}$ in the following subsections.

One can see in equations~(\ref{eq:timeL}) and (\ref{eq:timeS}) 
that following processes are implemented in our model: 
stars produce only large grains; 
shattering reduces large grains and increases small grains; 
coagulation reduces small grains and increases large grains; 
accretion increases only small grains; 
and SN shocks destroys both large and small grains.
As we describe further in the following subsections, accretion and coagulation 
occurs only in dense ISM, and shattering occurs only in lower density ISM. 
Accretion is the only process that genuinely increases the dust amount, 
and shattering and coagulation do not change the total mass of dust. 
One further important point in our formulation is that `astration', 
i.e., the consumption of dust (as well as gas and metals) by star formation
is automatically taken care of in our computation, as our dust formulation 
is using dust-to-gas ratio instead of dust mass itself.  
If the amount of gas is reduced by star formation, the amount of dust on 
the relevant gas particle is also reduced by the same factor. 
The dust in stars is assumed to be locked and does not change as a function of time 
after the formation of star particles. 

The current setup of the code does not allow us to treat delayed
dust input from AGB stars. However, contribution from AGB stars
can be effectively included in $f_\mathrm{in}$, although AGB
star dust production occurs at the same time as SN dust production
in our treatment. As compiled by \cite{2011MNRAS.415.2920I} and \cite{2013MNRAS.436.1238K},
the value of $f_\mathrm{in}$ is in the range of $\sim 0.01 - 0.5$, 
and it varies among theoretical models adopted. 
Since implementing delayed metal input requires a deeper modification
of the code, we only vary $f_\mathrm{in}$ to examine the effect of
dust production by stars, given that the uncertainty in $f_\mathrm{in}$
is large. Moreover, as shown later, interstellar processing, especially
dust mass growth by accretion, is much more efficient than stellar
dust input at ages $\ga$ several hundreds Myr, when AGB stars
would start to contribute \citep[][]{2009MNRAS.397.1661V}.
We leave delayed dust input from AGB stars to future work.


\subsection{Dust destruction in SN shocks}
\label{SNdestruction}

Each SN destroys the dust in its sweeping radius.
Since each SN is not spatially resolved in our simulations,
we simply adopt the following analytic formula for swept gas mass $m_{\rm sw}$ 
from \citet[][]{1989IAUS..135..431M}:
\begin{eqnarray} 
m_{\rm sw}&=&6800M_{\odot}
\left(\dfrac{E_{\rm SN}^{(1)}}{10^{51}~{\rm erg}}\right)
\left(\dfrac{v_{\rm s}}{100~{\rm km}~{\rm s}^{-1}}\right)^{-2}~,
\label{eq:m_sw}
\end{eqnarray}
where $E_{\rm SN}^{(1)}$ is the energy of a single SN 
(we fix it to $E_{\rm SN}^{(1)}=10^{51}$ erg).  
The shock velocity $v_{\rm s}$ is adopted from  
\citet[][]{1987ApJ...318..674M}, 
which is based on the Sedov-Taylor solution for single SNe:
\begin{eqnarray} 
v_{\rm s}&=&200\left( n_{0} \slash 1~{\rm cm}^{-3} \right)^{1/7}
\left( E_{\rm SN}^{(1)} \slash 10^{51}~{\rm erg} \right)^{1/14} {\rm km}~{\rm s}^{-1},
\end{eqnarray}
where $n_{0}$ is the number density of ambient gas. 

A single SN destroys the dust in each gas particle by the following fraction:
\begin{eqnarray} 
\eta&=&
\begin{cases}
\varepsilon_{\rm SN}\left(\dfrac{m_{\rm sw}}{m_{\rm g}} \right)
(m_{\rm sw} < m_{\rm g})~,\\
\varepsilon_{\rm SN}~(m_{\rm sw} \ge m_{\rm g})~,
\end{cases}
\end{eqnarray}
where $\varepsilon_{\rm SN}$ is the efficiency of dust destruction
in an SN blast wave. We adopt $\varepsilon_{\rm SN} = 0.1 $ 
based on \cite{2006ApJ...648..435N} and \cite{1989IAUS..135..431M}
unless otherwise stated \citep[see also][]{1994ApJ...433..797J,1996ApJ...469..740J}.

We estimate the number of SNe affecting the $i$-th particle, $N_i$, as
\begin{eqnarray} 
N_{i}&=&{\rm floor}\left(
\dfrac{E_{\rm SN}}{E_{\rm SN}^{(1)}}W 
\left(|\boldsymbol{r}_{i}|,h^{\rm exp}_{\ast}
\right)\right)~,\label{floor}
\end{eqnarray}
where ${\rm floor}(x)$ is Gauss' floor function
that returns the integer part of the argument, and 
${E_{\rm SN}}\slash{E_{\rm SN}^{(1)}}$ is the number of SN explosions.
Since the survival fraction of the dust is $1-\eta$ after an SN, 
the reduction of dust abundance after $N$ SNe can be estimated as
\begin{eqnarray} 
{\Delta \mathcal{D}_{{(\rm SNe\slash L, S)}(i)} (t)}=
\left[ 1-( 1 - \eta )^{N}\right] {\mathcal{D}_{{(\rm SNe\slash L, S)}(i)}(t)}, 
\label{eq:dustdestruct-1}
\end{eqnarray}
which is used in equations \eqref{eq:timeL} and \eqref{eq:timeS}.

\subsection{Subgrid models for dust physics}
\label{subgrid_model}
Accretion and coagulation of dust occur in dense molecular clouds
\citep{2014MNRAS.437.1636H}, which cannot be fully resolved by our current SPH simulations. 
Therefore, we have to treat these processes by subgrid models as we describe below. 

We impose following criteria for the selection of gas particles 
hosting unresolved dense gas:
\begin{eqnarray}
n_{\rm gas}> 10~{\rm cm}^{-3} \quad {\rm and}\quad T_{\rm gas}< 10^{3}~{\rm K}. 
\end{eqnarray}
We call these gas particles {\it dense gas particles}, and they occupy lower right 
corner of the phase diagram shown in Fig.~\ref{fig:phase}. 
We also assume that 
the fraction $f_{\rm dense}$ of each dense gas particle
is dense enough to allow accretion and coagulation, and that
the hydrogen number density and temperature of such dense media are
$10^{3}~{\rm cm}^{-3}$ and 50\,K, respectively.
We adopt $f_{\rm dense}=0.5$ as a fiducial value, 
and examine the effect of varying $f_{\rm dense}$ in Section~\ref{sec:f_dense}.

\subsubsection{Grain Growth by Accretion}
Accretion is a process in which small grains gain their mass 
by accreting gas from ambient medium. 
We obtain the time-scale of accretion from equation\, 23
in \citet{2011MNRAS.416.1340H}:
\begin{eqnarray} 
\tau_{\rm acc}=\begin{cases}
1.2\times 10^{6}{\rm ~yr}\left(\dfrac{Z}{Z_{\odot}} \right)^{-1}\left(1-\dfrac{\mathcal{D}_{\rm tot}}{Z}\right)^{-1}\slash f_{\rm dense}\\
\hspace{3cm}\mbox{(for \textit{dense gas particles})},\\
\infty~\mbox{(otherwise)}.
\end{cases}\label{accretion1}
\end{eqnarray}
In deriving this equation, we adopted
hydrogen number density $n_\mathrm{H}=10^3$ cm$^{-3}$,
gas temperature $T_\mathrm{gas}=50$ K, sticking efficiency of
gas-phase metals $S=0.3$
and typical radius of small grains
$a=5\times 10^{-3}~\micron$.
The infinity for $\tau$ means that we turn off the process.
The term 
$\left(1-{\mathcal{D}_{\rm tot}}\slash{Z}\right)^{-1}$
expresses the fact that only the metals in the gas phase
contribute to dust growth.
We note that the above time-scale is one-third of that 
in \cite{2012MNRAS.424L..34K}, 
who treated the time-scale in terms of grain radius,
while we define the time-scale by mass growth rate.

\textcolor{black}{{
\citet[][]{2016arXiv160804781Z} showed that the dust-to-metal ratio (depletion) is overestimated if a perfect sticking of metals on to dust ($S = 1$) is assumed. Moreover, they also argued that a decreasing sticking efficiency with gas temperature is required to reproduce the observed density dependence of depletion in the Milky Way ISM. Although our simulation is not capable of resolving clouds hosting dust growth, our implementation is consistent with \citet[][]{2016arXiv160804781Z}'s conclusion in that we only constrain dust growth to low-temperature regions and adopt a conservative sticking efficiency (0.3). The change of $f_\mathrm{dense}$ is also degenerate with that of $S$. We will later examine a case of a smaller $f_\mathrm{dense}$, which has the same effect as a lower $S$ (Section \ref{sec:f_dense}).}}

\subsubsection{Coagulation}
Coagulation is a process in which small grains collide with each other 
and turn into large grains. 
Since it is a collisional process between dust grains,
its time-scale $\tau_{\rm co}$ can be estimated 
based on the collision time-scale (see Appendix~\ref{collisional}):
\begin{eqnarray} 
\tau_{\rm co}=
\begin{cases}
2.71 \times 10^{5}{\rm ~ yr}\left( \dfrac{\mathcal{D}_{\rm S}}{0.01} \right)^{-1} 
\left( \dfrac{v_{\rm co}}{0.1 {\rm ~km}{\rm ~s}^{-1}} \right)^{-1} \slash f_{\rm dense}\\
~~~~~~~~~~~~~~~~~~~~~~~~~~~~~~~~~~~~({\rm for~{\it dense~clouds}})~,\\
\infty~({\rm for~others})~.\label{coagulation1}
\end{cases}
\end{eqnarray}
In using equation (\ref{eq:tau_coll}), we adopted
$\mathcal{D}=\mathcal{D}_\mathrm{S}$,
$a = 5\times 10^{-3}~\micron$ for the typical radius of small grains,
$s = 3$ g cm$^{-3}$ for the grain material density
and $v = v_{\rm co}$ for the typical velocity dispersion of small grains. 
We adopt $v_{\rm co}=0.1$ km s$^{-1}$ for the fiducial run.
Here, the typical velocity dispersion is based on 
the turbulence-induced grain velocity dispersion
calculated by a magnetohydrodynamic turbulence model in cold dense molecular clouds
\citep[][]{2004ApJ...616..895Y}.
Since the uncertainty in the velocity dispersion is large,
we also examine the case for $v_{\rm co}=0.01$ km s$^{-1}$
in Section~\ref{discussion1}.

\subsection{Shattering}
Shattering is a process in which large grains collide with each other
and get shattered into small grains. 
Since shattering occurs in the diffuse ISM \citep[][]{2009MNRAS.394.1061H},
it can be spatially resolved in our simulations, and no subgrid model is 
necessary. We assume that shattering occurs only 
in gas particles whose gas density is lower than 
the shattering threshold density $n_{\rm th}^{\rm SH}$.
We estimate the shattering time-scale 
based on the collisional time-scale in Appendix \ref{collisional}
using directly the gas density of gas particles as 
\begin{eqnarray} 
\tau_{\rm sh}&=&
\begin{cases}
5.41 \times 10^{7}~{\rm yr}
\left(\dfrac{\mathcal{D}_{\rm L}}{0.01}\right)^{-1}
\left(\dfrac{n_{\rm gas}}{1~\mathrm{cm}^{-3}}\right)^{-1}\\
\hspace{3cm}(n_{\rm gas} < n_{\rm th}^{\rm SH})~,                  \\
\infty~(n_{\rm gas} \ge n_{\rm th}^{\rm SH} )~.\label{shattering2}
\end{cases}
\end{eqnarray}
In estimating equation (\ref{shattering2}), we adopted
$a=0.1~\micron$, $s=3$ g cm$^{-3}$, $v=10$ km s$^{-1}$
(a typical velocity dispersion of large grains; \citealt{2004ApJ...616..895Y}),
$\mathcal{D}=\mathcal{D}_\mathrm{L}$, and
$n_\mathrm{H}=n_\mathrm{gas}$. 
We adopt $n_{\rm th}^{\rm SH} = 1$ cm$^{-3}$ 
unless otherwise stated.


\section{Results}

\subsection{Dust enrichment}\label{subsec:enrichment}
\begin{figure}
\includegraphics[width=8.5cm]{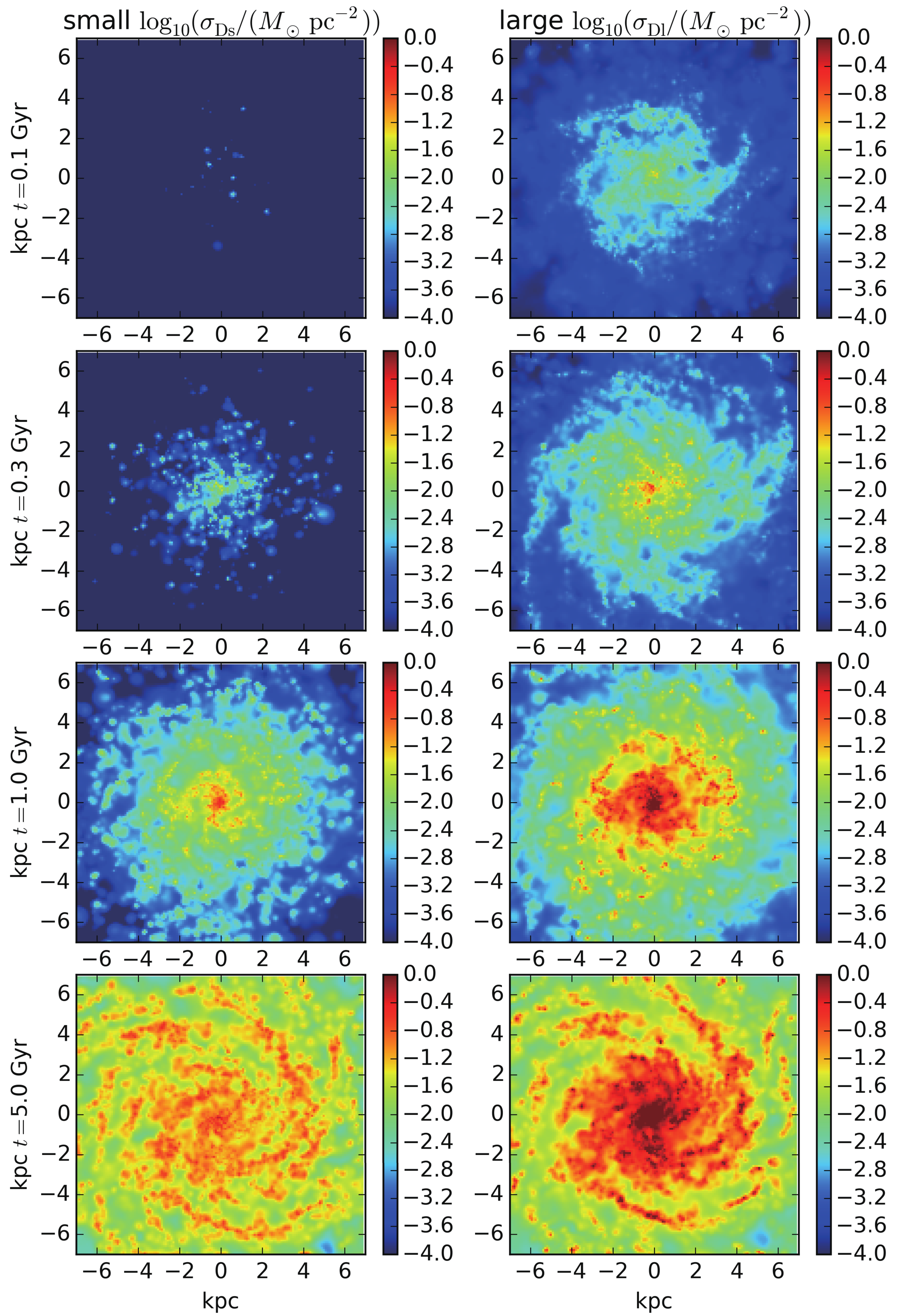}
\caption{
Face-on view of the surface densities of 
small grains (left-hand column) and large grains (right-hand column)
at $t=0.1, 0.3, 1,$ and 5 Gyr ({top to bottom}).}
\label{fig:dust_surface}
\end{figure}

One of the most fundamental features in our calculation is the
spatial distribution of dust in a galaxy.
We present the time evolution of surface densities of large and small grains 
at $t=0.1$, 0.3, 1, and 5 Gyr in Fig.~\ref{fig:dust_surface}.

At $t\simeq 0.1$ Gyr, the dust-abundant region is limited to
the central region of the galaxy because
of the relatively high SFR (i.e.,\ high
dust formation by stellar sources).
The processes of increasing the small grain abundance 
(shattering and accretion) are not efficient yet because of
low dust abundance. Thus,
the amount of small grains is much smaller than
that of large grains in the early phase of evolution.

At $t\simeq 0.3$ Gyr, the distribution of large grains
becomes more similar to the gas distribution as
the dust enrichment becomes more prevalent in the entire disc.
As time passes, 
the existence of small grains becomes clearer and clearer
because large grains are converted to small ones 
via shattering.
After $t\simeq 1.0$ Gyr, accretion becomes dominant
(Section~\ref{subsec:contribution}) and the
small grain abundance becomes comparable to the large grain abundance
in the inner region of the galaxy.
This is because the high-density environment in the central part raises 
both accretion and shattering efficiencies.

At $t\simeq 5.0$ Gyr, not only large grains but also small grains 
exist in a wide area of several kpc from the galactic centre, tracing the gas distribution well.
The density contrast becomes higher than that at $t=0.1$ Gyr
because of the non-linear dependence of accretion on density.
In addition, large grains are formed not only via stellar production 
but also via coagulation in this phase, 
because the coagulation rate becomes sufficiently high 
as the small grain abundance increases at later times (Section~\ref{subsec:contribution}).

\subsection{Comparison between large and small grain abundances}

\begin{figure*}
\includegraphics[width=14.0cm]{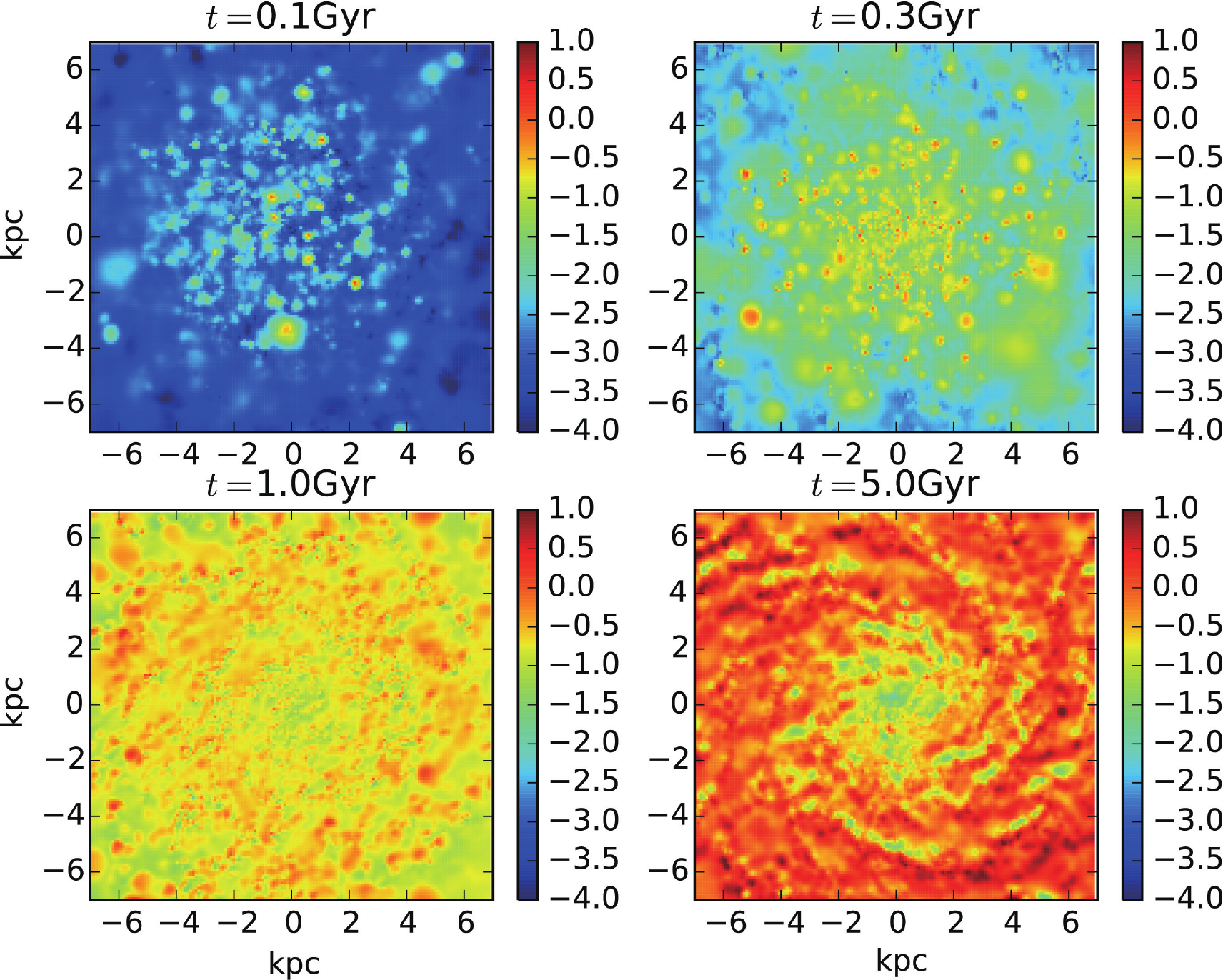}
\caption{Time evolution of 
$\log_{10}\left( \mathcal{D}_{\rm S} \slash \mathcal{D}_{\rm L} \right)$ 
at $t = 0.1, 0.3, 1$ and 5 Gyr.
Deep blue colour is used for $\mathcal{D}_{\rm S}\sim 0$, 
corresponding to $-4$ dex on the colour bar. 
}
\label{fig:ratio_surface}
\end{figure*}

We have newly incorporated the two-size approximation, which
provides information on grain size distribution. Thus, we are able to
show the spatially resolved grain size distribution map in a galaxy for
the first time, although the grain size distribution is represented by
two sizes to save the computational time.
In Fig.\ \ref{fig:ratio_surface}, we present
the surface density ratio of small to large grains, $\log_{10}\left( \mathcal{D}_{\rm S} \slash \mathcal{D}_{\rm L} \right)$. 
In the early stage at $t=0.1$ Gyr,  as discussed in Section \ref{subsec:enrichment},
the abundance of large grains is dominant over that of small grains
(i.e. most of the area is covered by blue or green colours)
in almost all regions of the galaxy, because the dust production is
dominated by stellar sources and $\mathcal{D}_{\rm L}$ is large. 

By comparing $ \mathcal{D}_{\rm S} \slash \mathcal{D}_{\rm L}$ at
various ages in Fig.~\ref{fig:ratio_surface}, 
we find a drastic increase of small grain abundance
between $t=0.1$ and 0.3 Gyr, as we already noted in
Section \ref{subsec:enrichment}.
Before $t=0.1$ Gyr, because the shattering time-scale is much shorter than 
that of accretion, small grains are generated predominantly by shattering.
After that, accretion is the dominant process for increasing small grain abundance
(Section~\ref{subsec:contribution}).
Because accretion is more efficient in more metal-rich and denser gas,
we find more regions with a high $\mathcal{D}_{\rm S}\slash \mathcal{D}_{\rm L}$
close to the central part than in the outskirts at $t\lesssim 0.3$ Gyr.
But in some localized dense regions, $\mathcal{D}_{\rm S}\slash \mathcal{D}_{\rm L}$
is already reaching close to unity, as seen by red spots of actively star-forming regions. 

At $t\simeq 1$ Gyr, shattering and coagulation start to show overall balance, 
and $\mathcal{D}_{\rm S} \slash \mathcal{D}_{\rm L}$ begin to reach 
a temporary convergence, and more wider regions start to show redder
colour approaching unity, with weaker dependence on the density.

After $t\simeq 5$ Gyr, $\mathcal{D}_{\rm S} \slash \mathcal{D}_{\rm L}$ 
reaches close to unity, and the outer part of galaxy also becomes largely 
red colour. 
In contrast, we observe a suppression of small grain abundance 
in the central part (green colour with $\mathcal{D}_{\rm S} \slash \mathcal{D}_{\rm L} \simeq 0.1$)
because of enhanced coagulation there.

\section{Discussion}\label{discussion1}

\subsection{Time evolution of total mass of each component}
\begin{figure}
\includegraphics[width= 8.0cm]{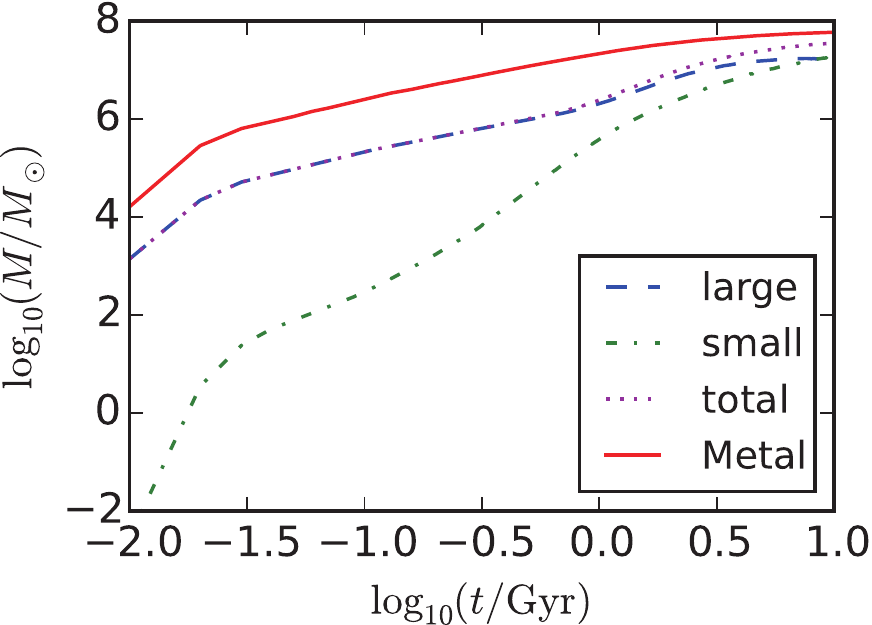}
\caption{Time evolution of total masses of metals (red solid), 
total dust mass (dotted), large grains (blue dashed), 
and small grains (dot--dashed). 
}
\label{fig:dustmass}
\end{figure}

We discuss the dust mass budget in the entire galaxy.
In Fig.~\ref{fig:dustmass}, 
we show the time evolution of the total masses of small and large grains.
We also present the total dust and metal masses 
(metals include both gas-phase and dust-phase components).
Because we assume that only large grains are created by stellar sources and 
adopt a fixed dust condensation efficiency $f_{\rm in}$, 
the mass ratio of large grains to metals is approximately $f_{\rm in} (= 0.1)$ 
in the early phase of the evolution. 
In addition, the abundance of small grains is much smaller 
in the early epoch such as $\lesssim 0.1$ Gyr.
This is because the small grains are created only via shattering and 
its time-scale is much longer than the age.

After $t\sim 0.1$ Gyr, small grains start to increase dramatically by shattering and accretion. 
However, the site where shattering and accretion occur efficiently is 
limited to the central part (Fig.~\ref{fig:dust_surface}) and 
the total abundance of dust is still much lower than that of metals.
At $\gtrsim 1$ Gyr, because small grains increase further by accretion,
a significant fraction of metals is locked into the dust.
The total mass ratio of small to large grains is 
(0.0036, 0.0659, 0.1795, 0.5641), 
and the total mass ratio of dust to metal is 
(0.0844, 0.1008, 0.3130, 0.6875) for 
$t=(0.1, 0.3, 1.0, 5.0)$ Gyr, respectively. 



\subsection{Contribution of each process}
\label{subsec:contribution}
\begin{figure*}
\includegraphics[width=16.0cm]{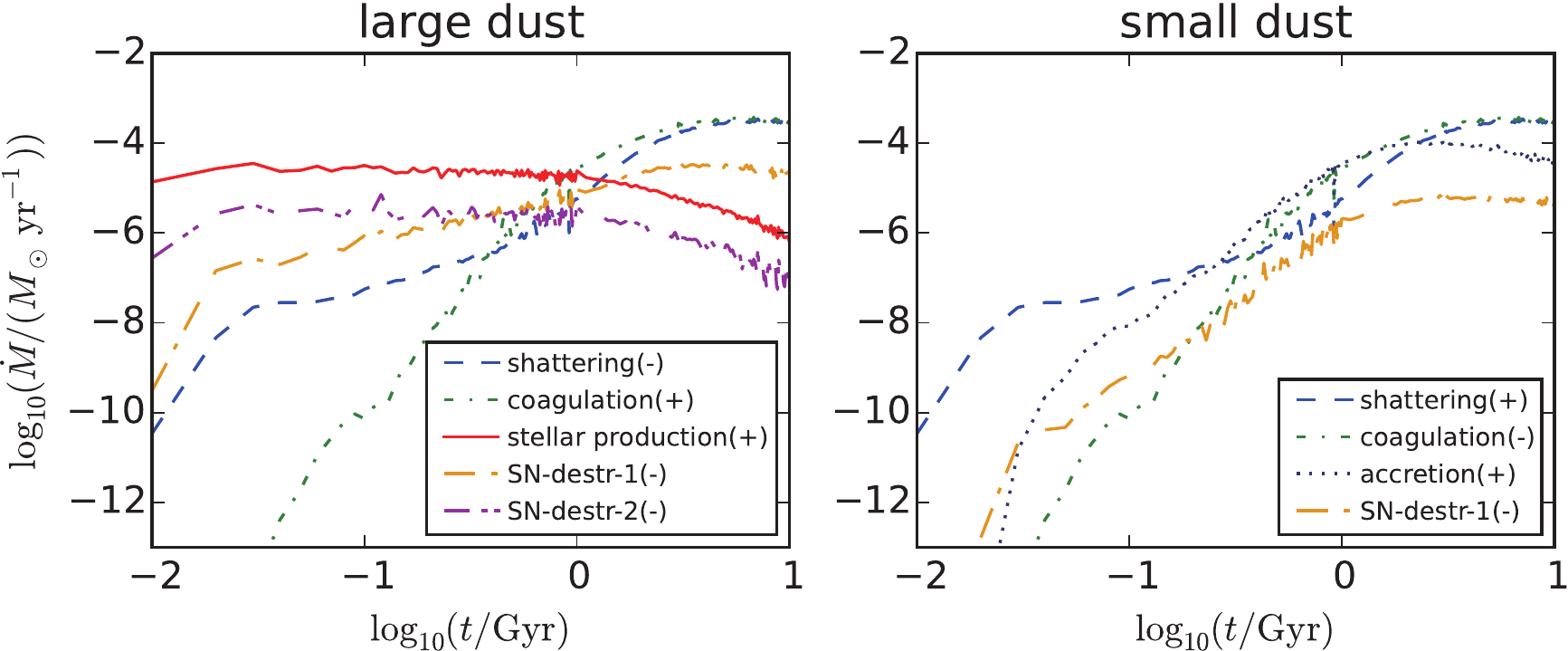}
\caption{Time evolution of dust production and destruction rate for each process. 
{Left-hand panel}: the terms from the right-hand side of equation~\eqref{eq:timeL} are shown for large grains: shattering (blue dashed), coagulation (green dot-dashed), stellar production (red solid), SN-destr-1 (orange dot-long-dashed; corresponding to the destruction rate of pre-existing dust in equation~\eqref{eq:dustdestruct-1}), and SN-destr-2 (purple dot-dot-dashed; corresponding to the destruction rate of newly formed dust,  $f_{\rm in } \dfrac{\Delta \tilde{m}_{\rm Z}}{m_{\rm g}} \delta$). 
{Right-hand panel}: the terms from the right-hand side of equation~\eqref{eq:timeS} are shown for small grains: 
shattering (blue dashed), coagulation (green dot-long-dashed), accretion (black dotted) and SN-destr-1 (orange dot-dashed; same as left-hand panel). 
The $\pm$ signs in the legend show the positive or negative impact of 
each process to the amount of large and small dust.
}
\label{dmdt}
\end{figure*}

In this section, we examine the contribution of each physical process
to dust production and destruction rate. 
In this calculation, we take into account all gas particles in the simulation, 
and sum up the contribution of each term on the right-hand side of 
Eq.~\eqref{eq:timeL} and \eqref{eq:timeS} for each process, 
as shown in Fig.~\ref{dmdt}.

Stellar dust production is the dominant production mechanism
of large grains in the early stage ($t\la 0.3$ Gyr).
For SN dust destruction,
we separate the destruction of newly formed dust ($f_{\rm in } \dfrac{\Delta \tilde{m}_{\rm Z}}{m_{\rm g}} \delta$ in equation~\ref{eq:timeL}) and
pre-existing dust (equation~\ref{eq:dustdestruct-1})
according to the formulation in equation~\eqref{eq:timeL}
and Appendix \ref{dustDestructionRate}. 
By construction of our model, 
the destruction rate of newly formed large grains (SN-destr-2 in Fig.~\ref{dmdt})
is proportional to the stellar dust production rate (red solid line in Fig.~\ref{dmdt}). 
Among all dust-processing mechanisms, at $t < 0.1$ Gyr, SN dust destruction 
of newly formed ones is the most dominant (SN-destr-2 in Fig.~\ref{dmdt}). 

On the other hand, the destruction rate
of pre-existing large grains (i.e., SN-destr-1 in Fig.~\ref{dmdt} and equation~\ref{eq:dustdestruct-1}) increases as the system is enriched with
dust, simply because the destruction rate is proportional to the destroyed material. 
The SN destruction rate of small grains also traces the increase of small grains.
There is no destruction of small newly formed grains, since we assume
that all grains formed by stellar sources are large.

The initial increase of small grains is governed by shattering
at $t\la 0.1$ Gyr. As the metallicity and dust abundance increase,
accretion catches up with the contribution from shattering,
becoming the dominant source of small grains around
$t\sim 0.1$ Gyr. As the small grain abundance increases further, 
the coagulation rate becomes comparable to 
the stellar dust production rate, which means that coagulation is the most
dominant mechanism of producing large grains.
Thus, the combination of accretion and coagulation is
important in increasing the dust abundance at
$t\ga 0.3$ Gyr (note that for the dust mass increase,
accretion is more fundamental than coagulation 
since coagulation itself does not increase the total dust mass).
Accretion saturates afterwards because a significant fraction
of metals are already accreted on the dust.
The contributions of coagulation and shattering become
comparable at $t\ga 1$ Gyr; thus, the small-to-large
grain abundance ratio is determined by the balance
between shattering and coagulation. Note that, since
these two processes do not change the total dust amount,
the total dust amount is still determined by the balance
between accretion and SN destruction.
The above evolutionary behaviours of various processes
are consistent with those in one-zone calculations 
\citep[][]{2013MNRAS.432..637A,2015MNRAS.447.2937H}.


\subsection{Dust abundance as a function of metallicity}

\begin{figure*}
\includegraphics[width=16.0cm]{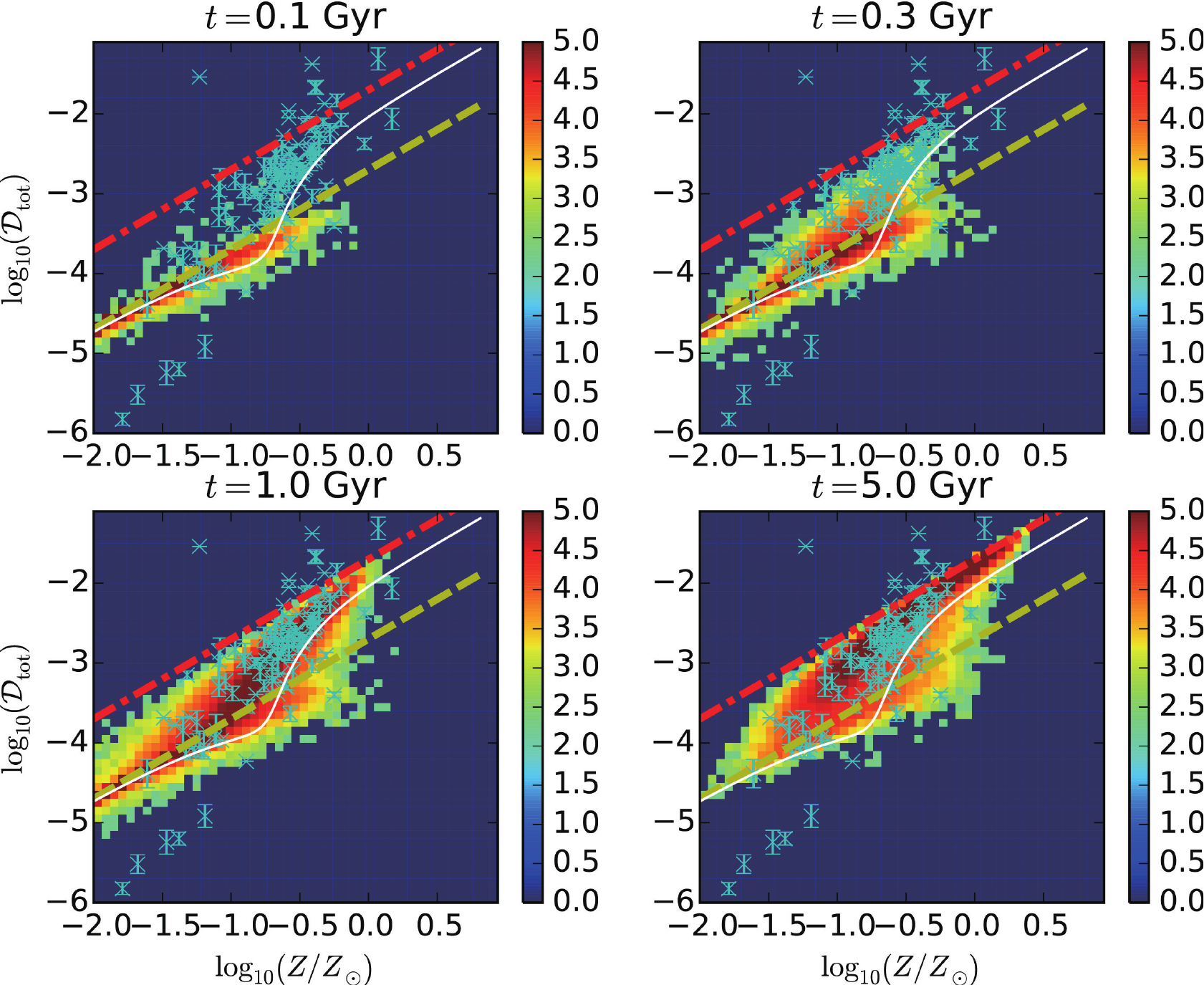}
\caption{Distribution of gas particles 
on the $\mathcal{D}_{\rm tot}-Z$ plane at 
$t=0.1, 0.3, 1$ and 5 Gyr as labelled. 
The colour indicates the logarithmic surface density of 
the gas particle on this diagram. 
The white line represents the one-zone calculation by \citet[][]{2015MNRAS.447.2937H}.
The yellow dashed and red dot-dashed lines denote the linear relation of the stellar yield $(\mathcal{D}_{\rm tot}=f_{\rm in}Z)$
and the saturation limit $(\mathcal{D}_{\rm tot}=Z)$, respectively.
Cyan crosses denote the observational data of 
nearby galaxies taken from \citet[][]{2014A&A...563A..31R} and
\citet[][]{2014A&A...562A..76Z}.
}
\label{fig_dust-gasMassRatio}
\end{figure*}

\begin{figure}
\includegraphics[width=8.3cm]{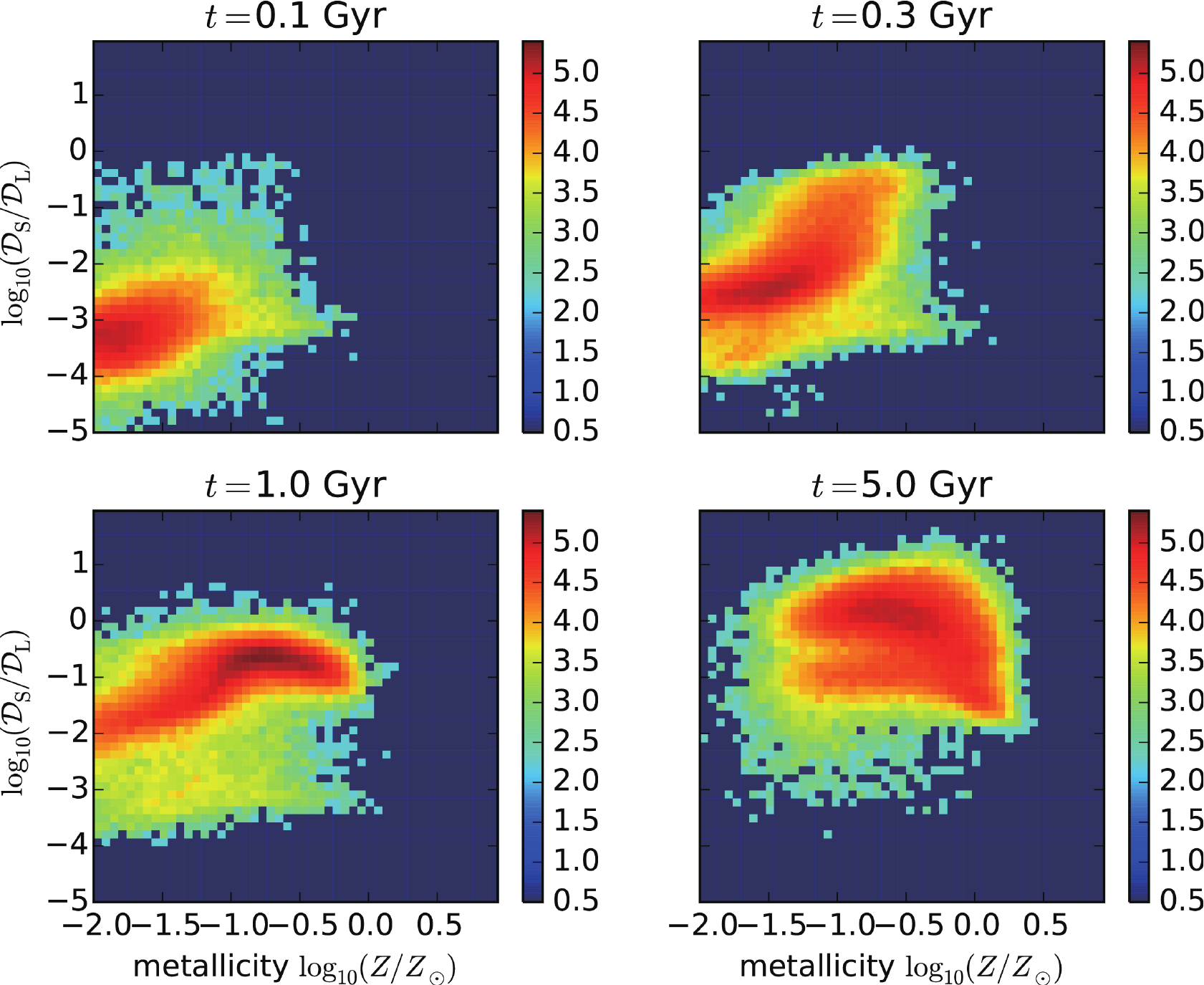}
\caption{Small-to-large grain abundance ratio ($\DSL$)
as a function of metallicity $Z$ at various epoch. 
The colour indicates the logarithmic surface density of the gas particle 
in each pixel of this diagram.}
\label{fig_large-smallRatio}
\end{figure}

We plot the time evolution of 
the relation between dust-to-gas ratio ($\mathcal{D}_\mathrm{tot}$)
and metallicity ($Z$) for all gas particles
in Fig.\ \ref{fig_dust-gasMassRatio}.

In this figure, we also plot the relation predicted by
the one-zone two-size approximation calculation by \citet{2015MNRAS.447.2937H} (white thin solid line). 
For the one-zone result, we choose their simple fiducial parameters for
various processes as a reference,
although in our simulations, the time-scale of each process varies from
particle to particle. We also show a simple 
constant dust-to-metal ratio expected from
pure stellar dust production $\mathcal{D}_{\rm tot}= f_{\rm in}Z$ (yellow dashed line) and 
the saturation limit $\mathcal{D}_{\rm tot}= Z$ (red dot--dashed line).

One can see that a fixed dust-to-metal ratio is 
a good approximation 
in the early phase ($t\la 0.1$ Gyr),  
since the dust evolution is driven by stellar production.
Many simulations simply assume a constant dust-to-metal ratio,
instead of solving the evolution of dust-to-gas ratio
\citep[e.g.,][]{2015MNRAS.451..418Y}. 
However, because accretion, whose efficiency responds to metallicity
nonlinearly, is dominating after that,
the fixed dust-to-metal ratio no longer gives a good approximation 
at $t\ga 0.3$ Gyr. The important feature of the
increase of dust-to-gas mass ratio due to accretion is that it
becomes prominent above a certain metallicity. This metallicity is called 
{\it critical metallicity} in, for example, \cite{2013EP&S...65..213A}.
This critical behaviour of accretion in terms of metallicity is caused
by the metallicity dependence of the accretion efficiency, as is clear
from the accretion time-scale in equation (\ref{accretion1}) 
\citep[see also][]{1998ApJ...501..643D,2008A&A...479..453Z}.
As the system is enriched with metals, the $\Dtot$--$Z$
relation extends towards higher metallicities and higher $\Dtot$.
Because accretion
saturates after a significant fraction of gas-phase metals are locked into dust,
the increase of $\Dtot$ as a function of $Z$
becomes moderate at the highest metallicity range.

We also plot the observed $\Dtot$--$Z$
relation for nearby galaxies with each point corresponding to
each individual galaxy 
\citep[not spatially resolved;][]{2014A&A...563A..31R,2014A&A...562A..76Z}. Since our
theoretical plots are for each SPH particle in our single-galaxy
simulation, strictly speaking, we cannot compare our results with
an unresolved nearby galaxy sample. Nevertheless, the fact that
our models roughly reproduce the observed $\Dtot$--$Z$
relation for nearby galaxies indicates that our implementation of
various dust formation and processing mechanisms is successful in
catching the trend of dust evolution as a function of metallicity.
\textcolor{black}{
Some previous works have also compared $\Dtot$--$Z$ relation obtained 
from their model calculations or simulations with observations 
\citep[][]{2003PASJ...55..901I, 2013MNRAS.432.2298B, 2016MNRAS.457.3775M, 2016arXiv160908622P}. 
For example, \citet[][]{2003PASJ...55..901I} argued that 
the $\Dtot$ increases rapidly by accretion when the galactic age is $\sim 0.3$\,Gyr.
Our simulation results are consistent with their results.  
\citet[][]{2016arXiv160908622P} found high dust-to-metal ratios in low-metallicity galaxies compared with the observation by \citet[][]{2014A&A...563A..31R}. 
In addition, the dust-to-metal ratio that they obtained is dependent on the model that they adopted. In our model, the ratio have a scatter 
due to hydrodynamics within a single galaxy. 
}

In Fig.~\ref{fig_large-smallRatio}, we also show the
relation between small-to-large grain abundance ratio and
metallicity, in order to clarify the evolution of grain size
distribution. In the earliest stage of evolution the main source
of small grains is shattering, whose efficiency increases in proportion
to the large grain abundance. While the large grain abundance increases
as a function of metallicity by stellar dust production,
the production rate of small grains depends on the square of
the large grain abundance.
Thus $\mathcal{D}_{\rm S}\slash \mathcal{D}_{\rm L}$ 
has a positive correlation with $Z$.
At $t \ga 0.3$ Gyr, the $\mathcal{D}_\mathrm{S}/\mathcal{D}_\mathrm{L}$
shows a strong dependence on $Z$ around $Z\sim 0.1$ $Z_{\sun}$ 
because of accretion.
At higher metallicities, as is clear in the plots at $t=1$ and 5 Gyr,
$\mathcal{D}_\mathrm{S}/\mathcal{D}_\mathrm{L}$ 
decreases as $Z$ increases because small grains turn into large ones via coagulation.
The above behaviour on the $\mathcal{D}_\mathrm{S}/\mathcal{D}_\mathrm{L}$--$Z$ 
diagram is consistent with the analysis in \citet{2015MNRAS.447.2937H}.

At $t=5$\,Gyr, the distribution of $\mathcal{D}_\mathrm{S}/\mathcal{D}_\mathrm{L}$ seems to
split into two components at $\log_{10}(Z / Z_{\odot}) \lesssim -0.5$.
The upper component with $\mathcal{D}_\mathrm{S}/\mathcal{D}_\mathrm{L}\sim 1$ corresponds to
low temperature ($T_{\rm gas}<10^{4}$ K) and diffuse gas ($10^{-2} \lesssim n_{\rm gas} < 1$ cm$^{-3}$) in the disc, and
the bottom component with $\mathcal{D}_\mathrm{S}/\mathcal{D}_\mathrm{L}\sim 0.1$ corresponds to
high temperature ($T_{\rm gas}>10^{4}$ K) and diffuse gas ($n_{\rm gas} < 1$ cm$^{-3}$, but largely $\lesssim 10^{-2}$ cm$^{-3}$;
see the phase diagram in Fig. \ref{fig:phase})
in both disc and circum galactic regions affected by SN feedback.
The former component has  higher densities than the latter component;
therefore the impact of shattering is much stronger, resulting in higher $\mathcal{D}_{\rm S}\slash \mathcal{D}_{\rm L}$.


\subsection{Parameter dependence}

There are some uncertain parameters involved in the above subgrid dust physics models. 
Although the values of those parameters are chosen based on physical reasoning, 
it is still worth examining the effect of varying them. 
Below, we examine how a representative parameter for each process affects the results. 
We use the
$\Dtot$--$Z$ and $\DSL$--$Z$ 
diagrams to examine the effects of various processes, since we have shown
above that these diagrams are useful for such a purpose.
We find that the qualitative behaviour as a function of time in this diagram 
does not change so much when we vary the parameters, 
therefore we focus on $t=1$ Gyr for this comparison. 

\subsubsection{Stellar dust production $f_{\rm in}$}
For stellar dust production, 
the condensation efficiency $f_{\rm in}$ is the key parameter. 
\textcolor{black}{\citet[][]{2008A&A...479..453Z} argued that 
the efficiency varies significantly depending on the SN type and elements.
}
\cite{2011EP&S...63.1027I} and \cite{2012MNRAS.424L..34K} found that 
$f_\mathrm{in}$ varies in the range of 0.01--0.5 
even among different theoretical calculations. 

\begin{figure}
\includegraphics[width=8.5cm]{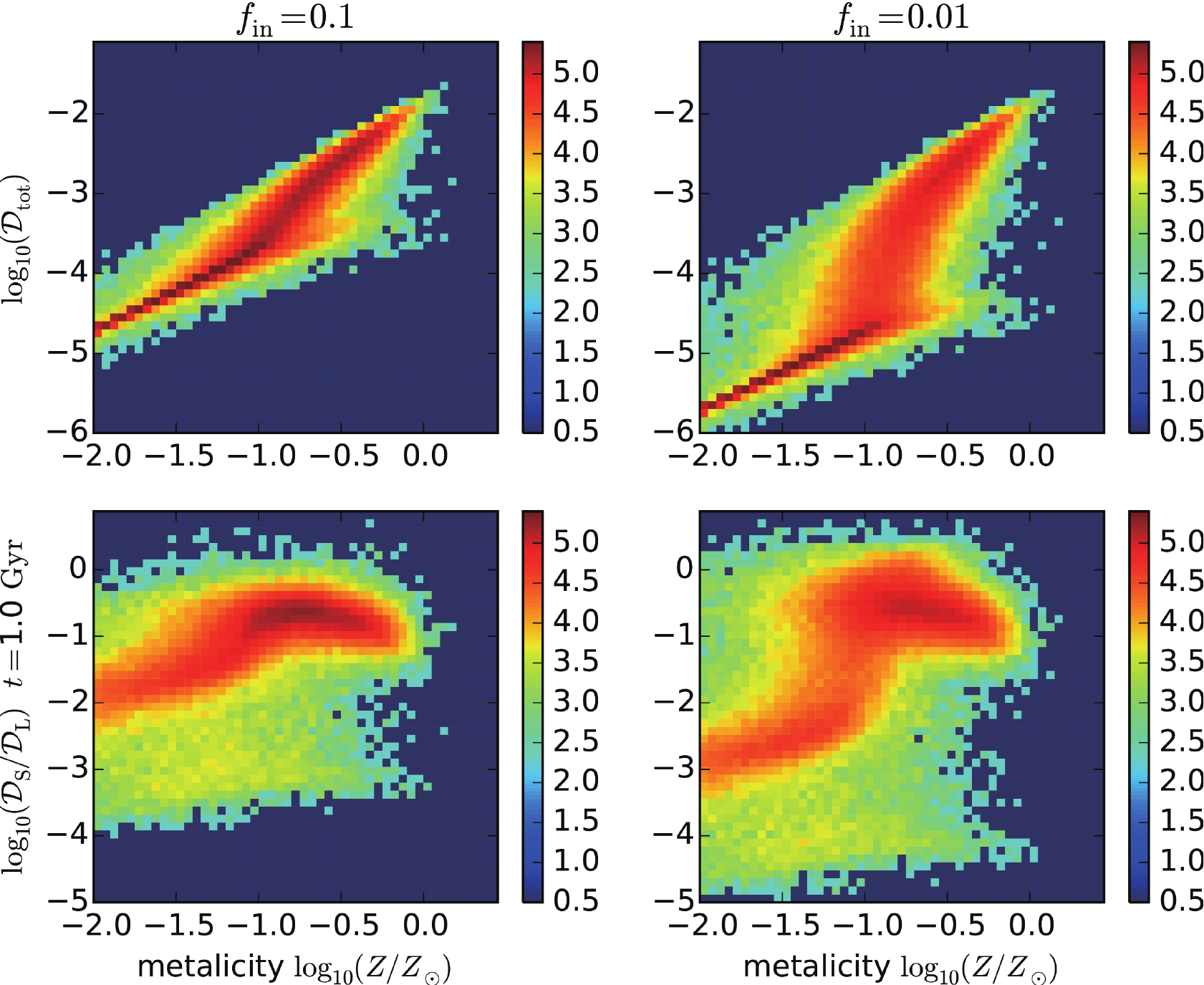}
\caption{Dependence of $\mathcal{D}_{\rm tot} $--$ Z$ (top row)
and $\mathcal{D}_{\rm S} \slash \mathcal{D}_{\rm L} - Z$ (bottom row)
relations on the dust coagulation efficiency $f_{\rm in}$ at $t=1$ Gyr.
In the left-hand column, we plot the distribution of gas particles 
for the fiducial case ($f_{\rm in} = 0.1$), which are also shown in 
Figs.~\ref{fig_dust-gasMassRatio} and \ref{fig_large-smallRatio}. 
In the right-hand column, we show the results of $f_{\rm in} = 0.01$ for comparison. 
}
\label{fig:f_in}
\end{figure}

To clarify the effect of varying $f_\mathrm{in}$,
we compare the results of $f_{\rm in}=0.01$ 
and our fiducial value $f_{\rm in}=0.1$ in Fig.~\ref{fig:f_in}.
The dust-to-gas ratio at low metallicity is simply proportional to $f_\mathrm{in}$ 
(i.e., $\mathcal{D}_{\rm tot}\sim \mathcal{D}_{\rm L}\sim f_{\rm in}Z$),  because
stellar dust production dominates the dust abundance increase. 
The value of $\mathcal{D}_\mathrm{S}/\mathcal{D}_\mathrm{L}$ at low metallicity is 
also lower for $f_\mathrm{in}=0.01$ than for $f_\mathrm{in}=0.1$,  
because the production rate of small grains 
by shattering is proportional to $\mathcal{D}_\mathrm{L}$ 
according to equation~\eqref{shattering2}.

However, both $\mathcal{D}_\mathrm{tot}$ and $\mathcal{D}_\mathrm{S}/\mathcal{D}_\mathrm{L}$
reach similar values at high metallicity regardless of the difference in $f_{\rm in}$,
because the processes other than stellar dust production, 
especially accretion, dominate these quantities.
Thus changing $f_{\rm in}$ affects the dust abundance only 
at low metallicities, typically $Z\la 0.1$ Z$_{\sun}$.

\subsubsection{Density threshold of shattering}

\begin{figure}
\includegraphics[width=8.5cm]{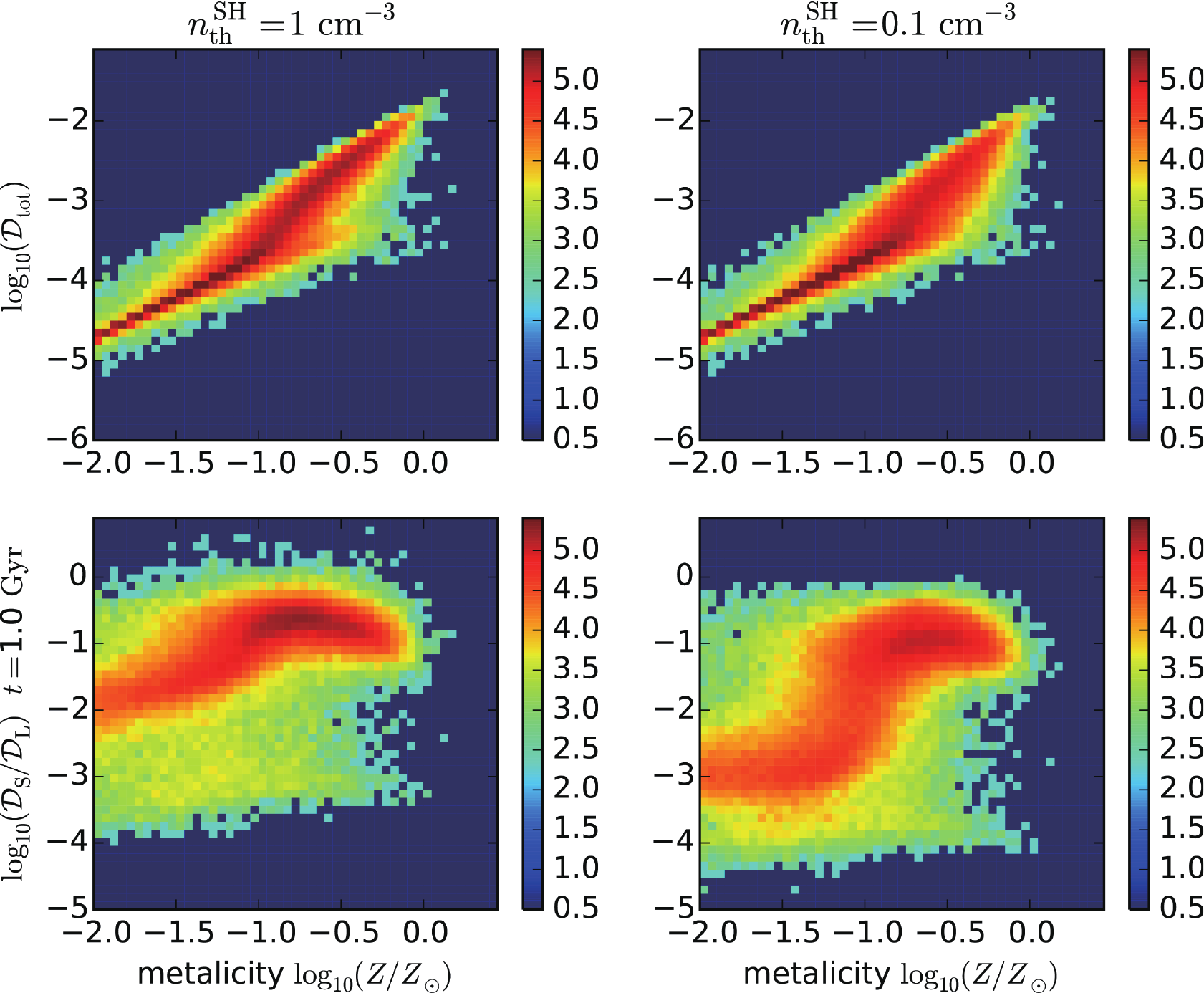}
\caption{Same as Fig.~\ref{fig:f_in},  
but for the comparison between 
$n_{\rm th}^{\rm SH} = 1$ cm$^{-3}$ (left-hand column)
and $n_{\rm th}^{\rm SH} = 0.1$ cm$^{-3}$ (right-hand column).
}
\label{fig:n_thSH}
\end{figure}

We also examine the dependence on the treatment of shattering. 
The strength of shattering is mostly regulated 
by the density threshold below which shattering occurs $(n_{\rm th}^{\rm SH})$.
Examining a low threshold density for shattering is especially important 
since setting too-high a threshold may cause 
an unreasonably strong shattering effect, 
because the shattering time-scale is inversely proportional to the density (equation \eqref{shattering2}). 
In Fig.~\ref{fig:n_thSH}, we compare the results for a lower density threshold 
$n_{\rm th}^{\rm SH} =0.1$ cm$^{-3}$ and the 
fiducial $n_{\rm th}^{\rm SH} =1$ cm$^{-3}$ (equation \ref{shattering2}).
Since shattering occurs in lower density gas than $n_{\rm th}^{\rm SH}$, 
lowering this threshold density reduces the amount of gas 
that will be affected by shattering. 

From this comparison, 
we find that the increase of $\mathcal{D}_{\rm tot}$ is 
slightly suppressed at $Z > 0.1 \Zsun$ for a lower $n_{\rm th}^{\rm SH}$, 
because less production of small grains by shattering leads to 
less efficient accretion.
Suppression of small-grain formation is clearly observed 
in lower $\mathcal{D}_{\rm S}\slash \mathcal{D}_{\rm L}$ at low metallicities
with lower $n_{\rm th}^{\rm SH}$ in Fig.~\ref{fig:n_thSH}, 
and the overall distribution of $\mathcal{D}_{\rm S}\slash \mathcal{D}_{\rm L}$ 
is also slightly lower even at higher metallicities.

\subsubsection{Grain velocity for coagulation}

The coagulation time-scale is regulated by the typical velocity of
small grains $v_\mathrm{co}$. 
As shown by \cite{2004ApJ...616..895Y}, the velocity of small grains 
in the dense medium depends strongly on gas density. 
Our fiducial value ($v_\mathrm{co}=0.1$ km s$^{-1}$) 
is taken from their calculation for molecular clouds,  
however, if we adopt their dense cloud cases, 
the velocity dispersion could be smaller by an order of magnitude. 
Thus, we examine the case of a smaller $v_\mathrm{co}=0.01$ km s$^{-1}$.

\begin{figure}
\includegraphics[width=8.5cm]{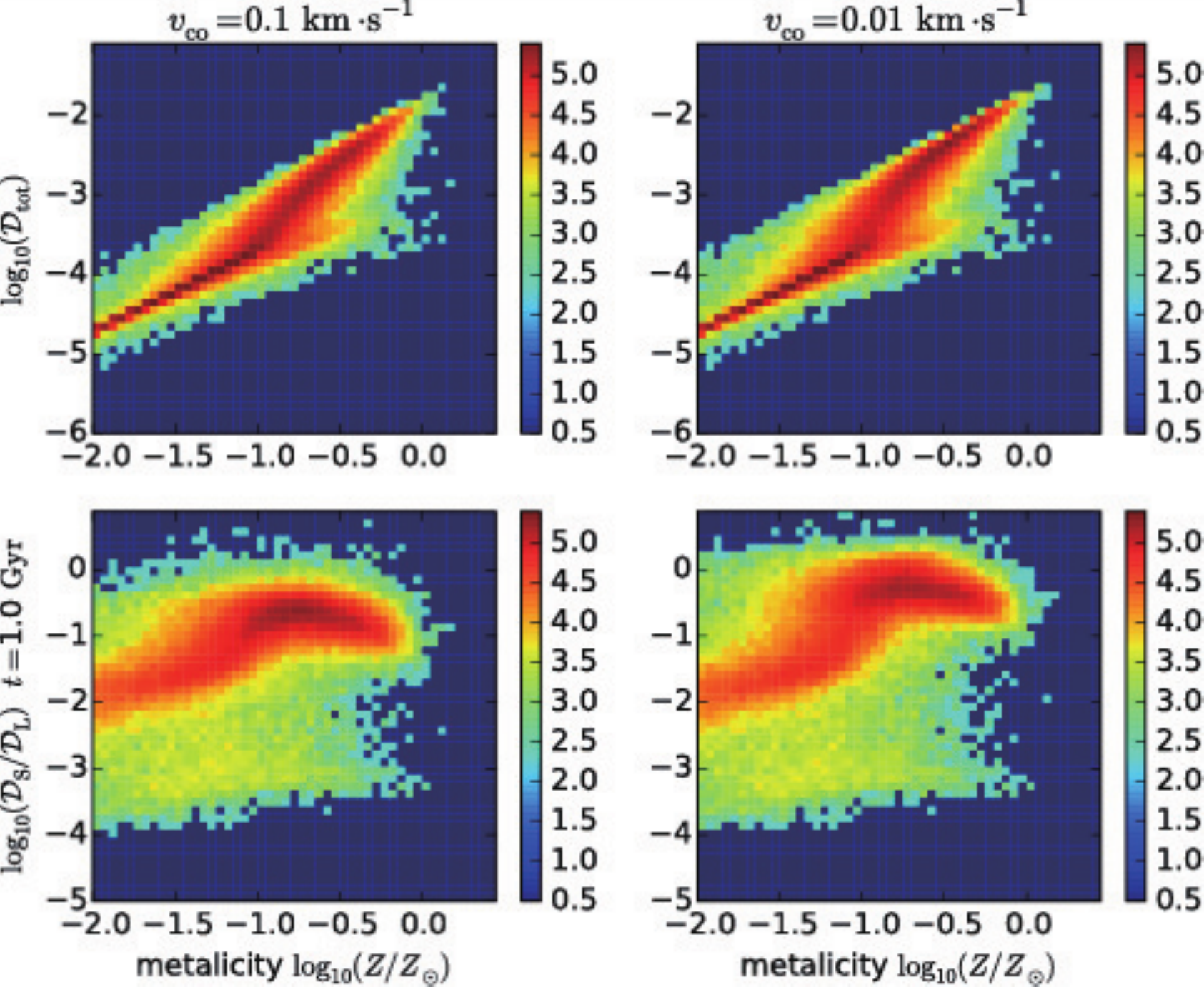}
\caption{Same as Fig.~\ref{fig:f_in}, but 
for the comparison between  $v_{\rm co}=0.1~{\rm km~s}^{-1}$ (left-hand column)
and $v_{\rm co}=0.01$ km s$^{-1}$ (right-hand column).
}
\label{fig:v_co}
\end{figure}

Fig. \ref{fig:v_co} compares the results with $v_\mathrm{co}=0.1$ (fiducial) and 0.01 km s$^{-1}$.  We observe that
a larger $v_\mathrm{co}$ (i.e., more efficient coagulation) predicts a lower $\DSL$ at high metallicities, as the small grains are converted into large grains via coagulation more. 
The resulting $\DSL$ is roughly 
inversely proportional to $v_{\rm co}$ at $Z \ga 0.3\, \Zsun$, 
because it scales with the inverse of coagulation efficiency.
Since coagulation conserves the total dust mass, 
$\Dtot$ is hardly affected by the change of $v_\mathrm{co}$.

\subsubsection{Mass fraction of dense cloud: $\fdense$}
\label{sec:f_dense}

\begin{figure}
\includegraphics[width=8.5cm]{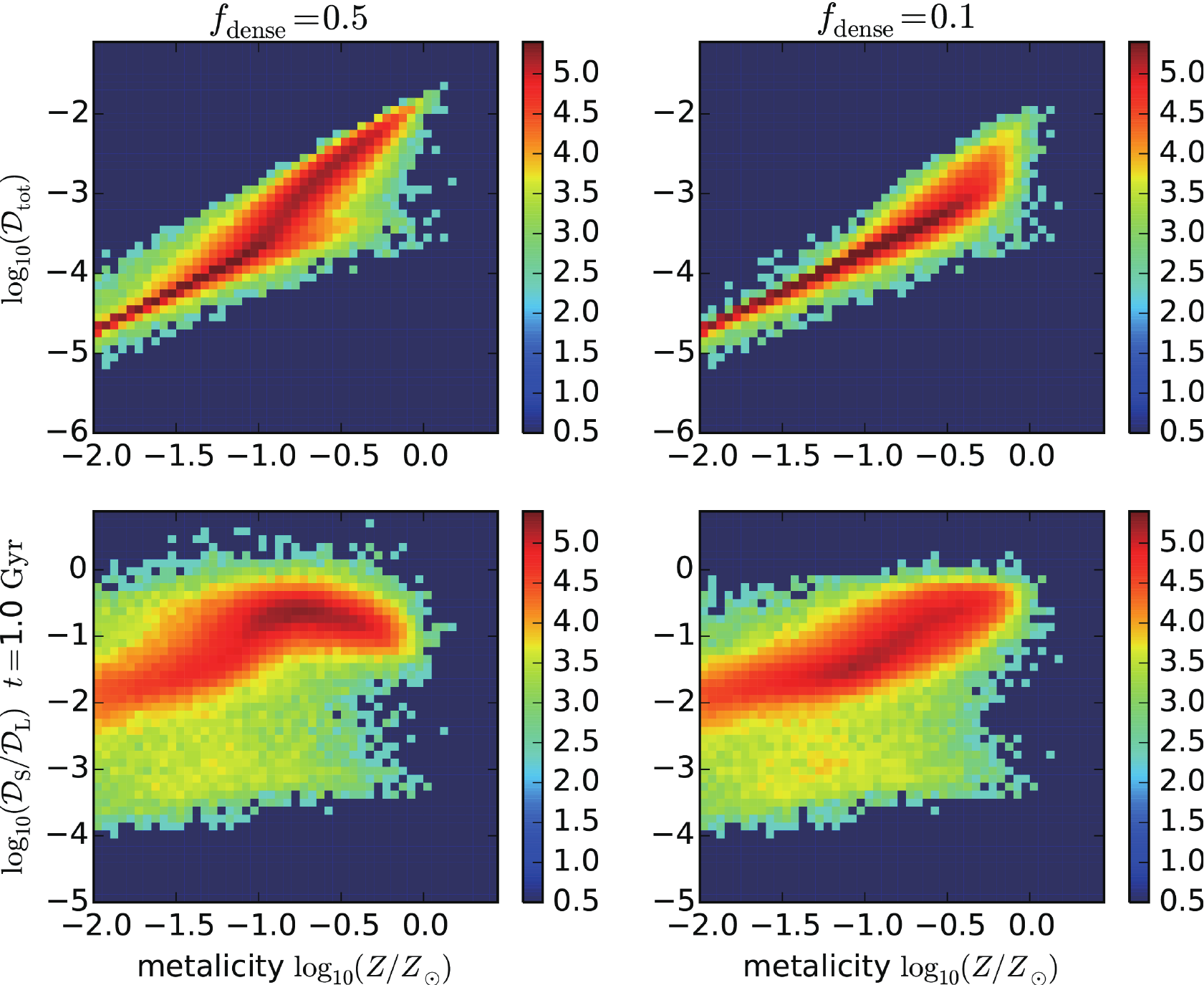}
\caption{
Same as Fig.~\ref{fig:f_in}, but for the comparison between 
$\fdense = 0.5$ (left-hand column) and $\fdense = 0.1$ (right-hand column).
}
\label{fig:v_f_dense}
\end{figure}

The subgrid parameter introduced to describe accretion 
and coagulation is the fraction of dense clouds $\fdense$ (Section~\ref{subgrid_model}). 
In addition to the fiducial case of $\fdense = 0.5$, 
we also examine a low fraction $\fdense =0.1$, as shown in Fig.~\ref{fig:v_f_dense}.

The increase of $\Dtot$ by accretion occurs 
at higher metallicity for $\fdense =0.1$ than for $\fdense =0.5$,  
because of a lower efficiency of accretion. 
In other words, a higher metallicity is required for accretion to be efficient enough 
to raise the dust-to-gas ratio. 
However, the dust-to-gas ratio eventually reaches similar values at solar metallicity
in both cases. 

The effect of coagulation is seen in $\DSL$, which is systematically lower at high metallicities
for $\fdense =0.5$ than for $\fdense =0.1$ due to more efficient coagulation. 
In contrast, the change of $\fdense$ does not affect the results at low metallicities, because neither coagulation nor accretion is the dominant mechanism for driving the dust evolution at low metallicities.

\begin{figure}
\includegraphics[width=8.5cm]{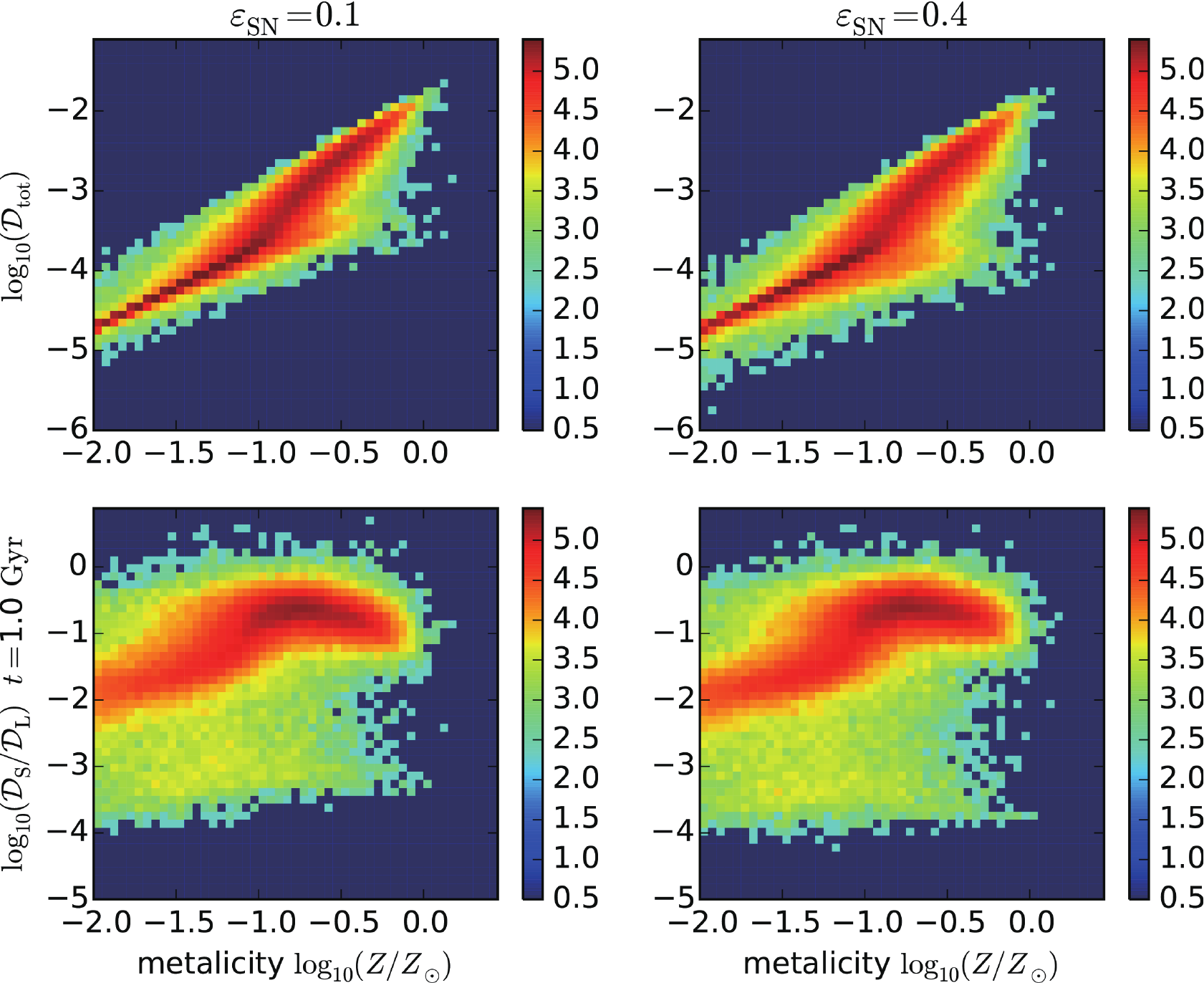}
\caption{Same as Fig.~\ref{fig:f_in}, but for the comparison between 
$\eSN = 0.1$ (left-hand column) and $\eSN = 0.4$ (right-hand column).
}
\label{fig:epsilon}
\end{figure}


\subsubsection{Dust destruction efficiency in an SN blast: $\eSN$}

\citet[][]{2006ApJ...648..435N} studied the efficiency of dust destruction 
in an SN blast for various dust species, and showed that 
it varies by about a factor of 4 among different dust species. 
Therefore we also simulate a high efficiency case 
with $\eSN = 0.4$ and compare with the fiducial case, $\eSN = 0.1$,
in Fig.~\ref{fig:epsilon}.

We find that the dispersion of gas particles on 
the $\Dtot$--$ Z$ plane at $Z\sim 0.1\, \Zsun$
is greater for $\eSN = 0.4 $ than for $\varepsilon _{\rm SN} = 0.1$.
The dispersion is especially increased towards small $\Dtot$. 
In addition, $\DSL$ is slightly suppressed at $Z\sim 0.03\, \Zsun$ 
for $\eSN=0.4$ case compared to $\eSN=0.1$ case.
These results suggest that the location of the increase of $\Dtot$ and 
$\DSL$ by accretion is shifted to higher metallicity if SN destruction is stronger. 
This is consistent with the one-zone calculation result 
in \citet[][]{2015MNRAS.447.2937H}.
The 'disturbance' by SN destruction also increases scatter at high metallicities.
Therefore, although SN destruction does not change 
the overall behaviour in the $\Dtot-Z$ diagram,
it produces more scatter in the dust abundance and grain size distribution.


\section{Observational implications}

Although the main purpose of this paper has been to establish a computational framework of 
dust evolution in galaxy simulations, 
it is an important step to compare the output of simulations with observational data. 
Among the quantities that can be compared directly with observations, 
we choose the radial distribution of dust-to-gas ratio and 
dust-to-metal ratio. 
The dust-to-metal ratio is referred to as the `depletion' in this paper.

Depletion is one of the observational quantities
that suggest the existence of dust in the ISM \citep[e.g.,][]{1990ARA&A..28...37M}.
\textcolor{black}{{A detailed depletion pattern for various elements is indeed derived for the Milky Way ISM by comparing a reference abundance with the observed ISM abundance \citep[e.g.,][]{2009ApJ...700.1299J}, although the spatial distribution (e.g., radial distribution) of depletion is difficult to infer in the Milky Way.}
}
There is no direct observational data for grain size distributions, 
although extinction curves and dust emission spectra 
may give some constraints on them
\citep[][]{2014MNRAS.439.3073Y, 2015MNRAS.447L..16N, 2016arXiv160806099H}. 
However, further modelling is necessary 
to predict dust extinction and emission, 
requiring additional sets of assumptions
\citep[][]{2001ApJ...551..807D,2001ApJ...554..778L}. 
Therefore we leave those issues for our future work (Hou et al., in preparation).


\subsection{Radial profile}

\begin{table}
{\footnotesize 
\begin{tabular}{lcccc}
Galaxy & $m_{\ast}$\,($10^{9}\Msun$) & $\psi$\,($\Msun\slash {\rm yr}$)
& sSFR\,(Gyr$^{-1}$) & Category  \\ \hline \hline
HolmbergII &$ 0.20 ^{(1)}$ &$ 6.61 ^{(1)}$ &$ 33.128 ^{(1)}$ & I \\ 
NGC925 &$ 7.94 ^{(1)}$ &$ 10.72 ^{(1)}$ &$ 1.350 ^{(1)}$ & II \\ 
NGC3621 &$ 58.88 ^{(2)}$ &$ 125.89 ^{(3)}$ &$ 2.138 $& II \\ 
NGC628 &$ 12.59 ^{(1)}$ &$ 9.77 ^{(1)}$ &$ 0.776 ^{(1)}$ & III \\ 
NGC2403 &$ 2.14 ^{(4)}$ &$ 1.74 ^{(4)}$ &$ 0.813 ^{(4)}$ & III \\ 
NGC4736 &$ 19.95 ^{(1)}$ &$ 7.76 ^{(1)}$ &$ 0.389 ^{(1)}$ & III \\ 
NGC5055 &$ 63.10 ^{(1)}$ &$ 11.75 ^{(1)}$ &$ 0.186 ^{(1)}$ & III \\ 
NGC5194 &$ 39.81 ^{(1)}$ &$ 7.76 ^{(1)}$ &$ 0.195 ^{(1)}$ & III \\ 
NGC7793 &$ 3.39 ^{(2)}$ &$ 3.16 ^{(3)}$ &$ 0.933 $& III \\ 
NGC2841 &$ 79.43 ^{(1)}$ &$ 6.92 ^{(1)}$ &$ 0.087 ^{(1)}$ &   IV \\ 
NGC3031 &$ 85.11 ^{(2)}$ &$ 12.59 ^{(3)}$ &$ 0.148 $     &   IV \\ 
NGC3198 &$ 10.00 ^{(4)}$ &$ 1.00 ^{(4)}$ &$ 0.100 ^{(4)}$ &   IV \\ 
NGC3351 &$ 56.23 ^{(4)}$ &$ 3.31 ^{(4)}$ &$ 0.059 ^{(4)}$ &   IV \\ 
NGC3521 &$ 190.55 ^{(4)}$ &$ 10.47 ^{(4)}$ &$ 0.055 ^{(4)}$ & IV  \\ 
NGC7331 &$ 79.43 ^{(1)}$ &$ 9.12 ^{(1)}$ &$ 0.115 ^{(1)}$ &   IV  \\ 
\hline
\end{tabular}
}\normalsize
\caption{Stellar mass $(m_{\ast})$, global SFR $(\psi)$, 
sSFR and 
corresponding category of each galaxy. 
The references of each values are described as a superscript.
(1), (2), (3) and (4) correspond to 
\citet[][]{2014AJ....147..103H},
\citet[][]{2013NewA...19...89D,2008AJ....136.2648D} and
\citet[][]{2015AJ....149....1Z}, respectively. 
For SFR and stellar mass, 
\citet[][]{2013NewA...19...89D} and \citet[][]{2008AJ....136.2648D} are used, respectively.
}\label{tab:sample}
\end{table}

\begin{figure}
\includegraphics[width=8.5cm]{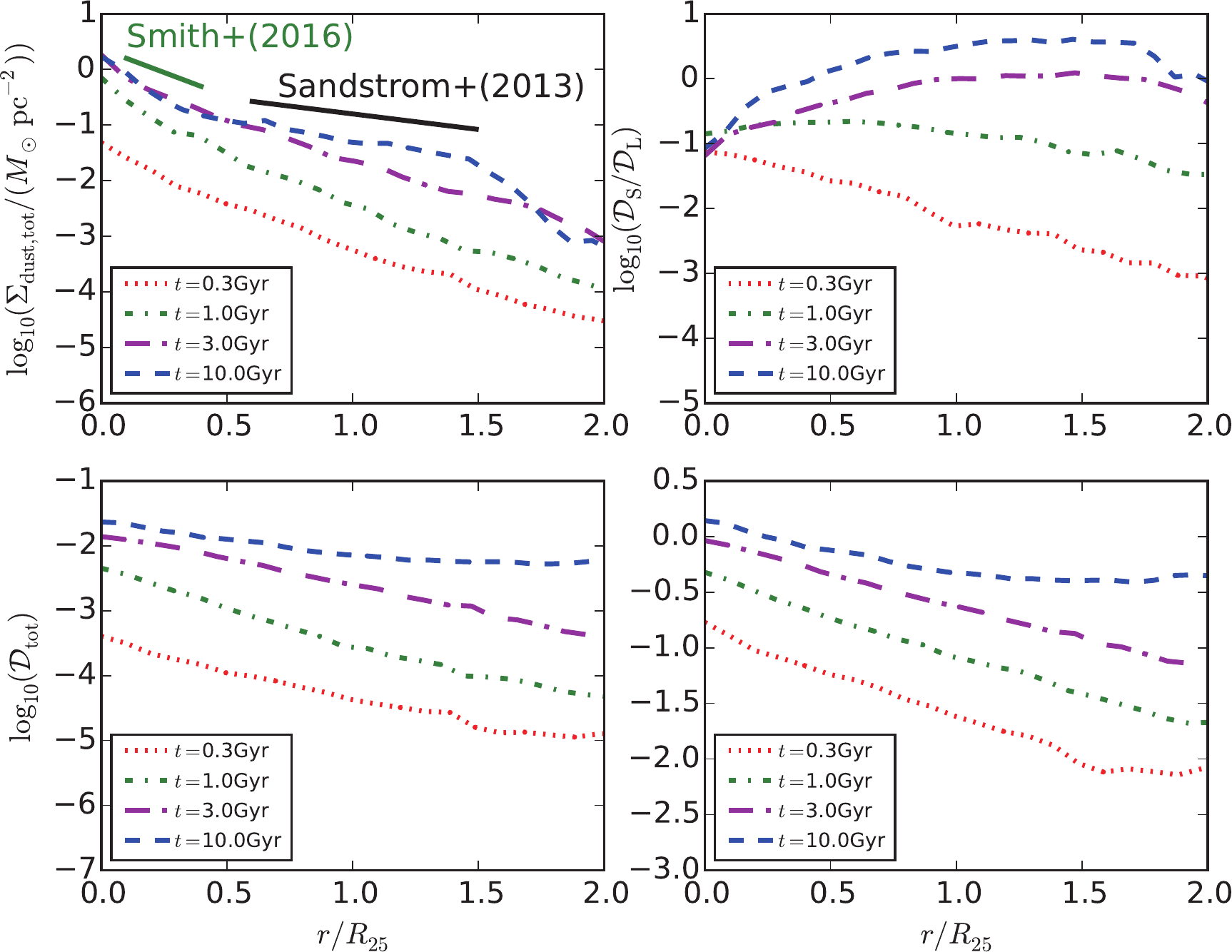}
\caption{
Radial profiles of the dust surface density $\Sigma_{\rm dust,tot}$ ({top-left}), 
small-to-large grain abundance ratio $\DSL$ ({top-right}), 
total dust-to-gas mass ratio $\Dtot$ ({bottom-left}), and 
metallicity $Z$ ({bottom-right}) at $t=0.3$ Gyr (dotted), 1.0 Gyr (dot-short-dashed), 
3.0 Gyr (dot-long-dashed), and 10.0 Gyr (dashed). 
In the top left panel, we also overplot the observationally derived radial profiles  
by \citet[][]{2013ApJ...777....5S} and \citet[][]{2016MNRAS.462..331S}. 
}
\label{fig:radialProfile}
\end{figure}

We discuss and compare the total dust surface density 
in simulations and observations. 
First, we show the time evolution of
the radial profile of dust surface density in Fig.~\ref{fig:radialProfile}. 
Overall, the dust surface density
increases as time passes due to dust enrichment.
The dust surface density rapidly increases from $t=0.3$ to 1\,Gyr due to accretion,
especially in the central region. 
However, the star formation activity in the central region is 
also high and the gas is consumed rapidly before $t \simeq 1$ Gyr.
Thus the dust surface density is saturated by astration (i.e., 
consumption of gas, metals and dust by star formation)
at the central region after $t \simeq 1$ Gyr. 
In contrast, the dust-to-gas ratio and metallicity monotonically increase at all radii 
even in this phase, because astration decreases
dust, metals and gas at the same rate.

The time evolution of radial profile of $\DSL$ can be understood in the following way.
In the early phase such as $t\lesssim 0.3$ Gyr, 
the source of small grains is mainly shattering and 
the abundance of small grains is determined only by that of large
grains. Hence $\DSL$ roughly follows the radial profile of $\Dtot$.
At $0.2 \lesssim t\lesssim 0.5$ Gyr,  small grains start to
increase more than large grains via accretion, especially in the
central region where the gas density (i.e., efficiency of shattering) 
and SFR (i.e., production rate of large grains) 
are high. The rapid increase in $\DSL$ is observed at later epochs
($\sim 1$ Gyr) in the outer region.  In contrast, in the central region,
coagulation decreases $\DSL$ at $t\ga 1$ Gyr.

In the later stage of simulation ($t\ga 1$ Gyr),
shattering further increases $\DSL$. 
The reason is the following. As time passes,
dense gas particles are consumed by  star formation and 
the sites of coagulation monotonically decrease.
On the other hand, the sites of shattering do not decrease 
so much, because diffuse ISM does not form stars.
As a result, $\DSL$ continues to increase after $t=1$ Gyr.

The radial profile of dust-to-gas ratio have been  
observationally investigated by \citet[][]{2013ApJ...777....5S}  for nearby galaxies.
To cancel the galaxy size effect, we normalize the radius by 
$R_{25}$ (the radius at which  surface brightness falls to 25 mag arcsec$^{-2}$)
for the simulated galaxy, following \cite{2012MNRAS.423...38M}.  
The value of $R_{25}$ changes as time passes, because the distribution of stellar component changes: 
$R_{25}=(6.9, 7.0, 7.2, 8.4)$ kpc at $t=(0.3, 1.0, 3.0, 10.0)$ Gyr, respectively.

The surface density profile of stellar disc in a spiral galaxy, $\Sigma_*(r)$, 
can be approximately described by a single exponential function as 
$\Sigma_*(r)=\Sigma_{0}\exp ( -{r}/{R_{\rm d}})$,
where $\Sigma_0$ is the central surface density. 
The scale length of stellar disc, $R_{\rm d}$, 
is determined by fitting the stellar surface density profile
of simulated galaxy at each epoch. 
We relate $R_\mathrm{d}$ and $R_{25}$ by $ R_{25} \simeq 4 R_{\rm d}$
 \citep[][]{1998ggs..book.....E}.
This relation can be confirmed by the data in, e.g., \cite{1978AJ.....83.1163D}.
\citet[][]{2013ApJ...777....5S} showed that 
the stacked radial profile of dust surface density is well explained by $\exp \left( -0.56\, r/R_{25} \right)$. 
We plot this slope in Fig.~\ref{fig:radialProfile} together with the simulation results, 
and find that the simulation results at $t>3.0$ Gyr agree well with the observed one
by \citet[][]{2013ApJ...777....5S}. 
\citet[][]{2016MNRAS.462..331S} also gave a fit of $-1.7$\,dex\,$R_{25}^{-1}$ to the 
observed dust surface density profile using 110 spiral galaxies in the {\it Herschel}
Reference Survey, as shown in the top-left panel of Fig.~\ref{fig:radialProfile}.

\begin{figure}
\includegraphics[width=7.7cm]{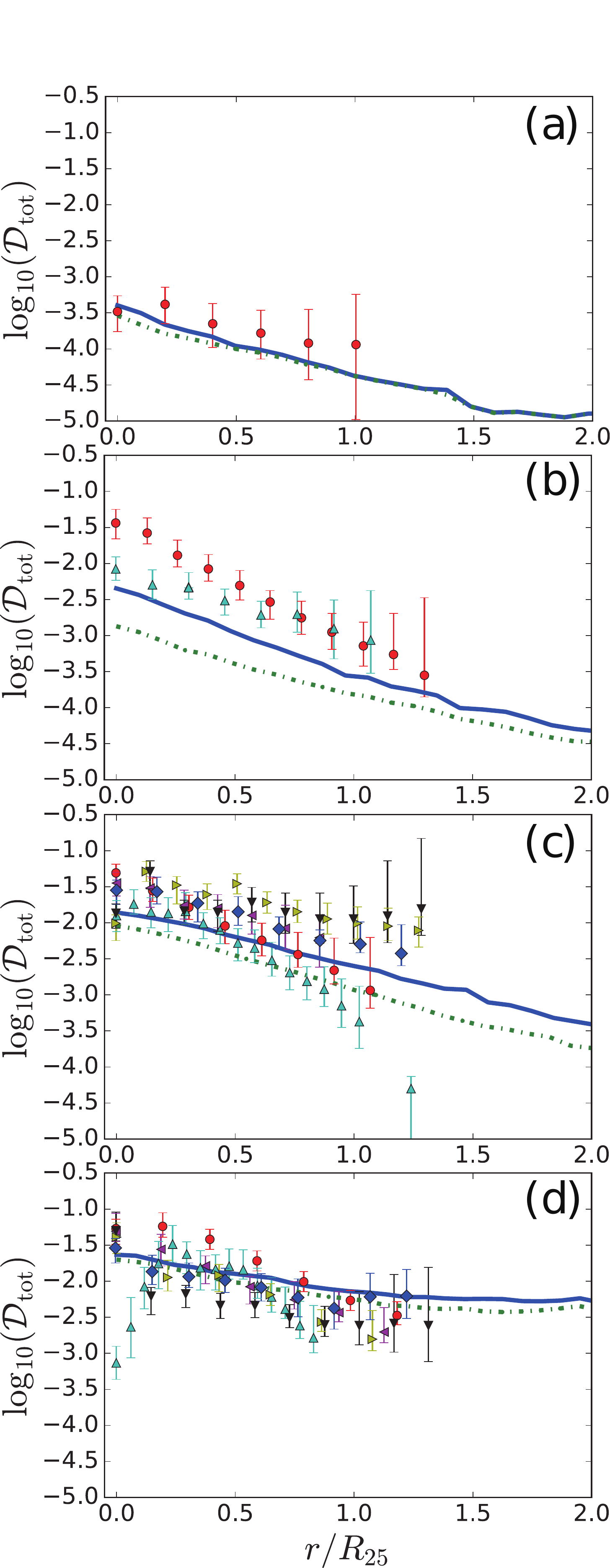}
\caption{Comparison of the radial profiles of total dust-to-gas ratio with 
observational data taken from \citet{2012MNRAS.423...38M}.
We scaled the radius with $R_{25}$.
In panel ({\it a}), circle points represent 
the observational radial profile of 
Holmberg II. 
In panel ({\it b}), circle and triangle points show
that of NGC 3621 and NGC 925, respectively.
In panel ({\it c}), circle, $\bigtriangleup $, $\lhd $, $\rhd $, $\bigtriangledown$
and diamond symbols correspond to NGC 628, NGC 2403, NGC 4736, 
\textcolor{black}{NGC5055, }
NGC 5194 and NGC 7793, respectively.
In panel (d), circle, $\bigtriangleup $, $\lhd $, $\rhd $, $\bigtriangledown$
and diamond symbols correspond to NGC 2841, NGC 3031, NGC 3198, NGC 3351, NGC 3521
and NGC 7331, respectively.
The solid and dotted lines represent the case with $f_{\rm dense } =0.5$ and 0.1, 
respectively.}
\label{fig:dust_tot}
\end{figure}

In addition, we also compare our simulation against the observed data by 
\cite{2012MNRAS.423...38M}, who investigated the radial profile of
dust-to-gas ratio and dust-to-metal ratio in nearby star-forming galaxies
(mainly spiral galaxies).  They used the metallicity calibration method 
of \citet[][]{2010ApJS..190..233M}.
In order to compare the simulation snapshots at various ages with observations,
we classify the observed galaxy sample according to their evolutionary stage.
We use the specific star formation rate (sSFR) as an indicator of evolutionary stage,
as the build-up of stellar mass is tightly related to the metal and
dust enrichment. We take sSFR in the literature as summarized in Table~\ref{tab:sample}, 
and divide the sample into following four categories: 
10 Gyr$^{-1} <$ sSFR,
1 Gyr$^{-1} \le $ sSFR $\le $ 10 Gyr$^{-1}$, 
$10^{-0.8}  < $ sSFR $ < 1 $ Gyr$^{-1}$
and sSFR $ < 10^{-0.8} $ Gyr$^{-1}$ for Category I, II, III and IV, 
respectively.
Holmberg II is a dwarf irregular galaxy, not a spiral galaxy,
but it is also included as a representative of young objects, 
although the comparison should be taken as preliminary 
until we really simulate a dwarf irregular galaxy.
In our simulated galaxy, we adopt the ages 0.3, 1, 3, and 10 Gyr, 
when the average sSFR is $\sim 3$ , 1, 0.3 and 0.1 Gyr$^{-1}$, respectively.
We compare the simulation results at
$t=0.3, 1, 3$ and 10 Gyr  with Category I, II, III and IV, respectively.

We plot the radial profile of $\Dtot$ of each category in Fig.~\ref{fig:dust_tot}.
Since the radial profile is sensitive to grain growth,
we show the results with $\fdense=0.5$ (solid line) and 0.1 (dotted line).
The dust-to-gas ratio in the `youngest' phase represented by Category I
(Holmberg II) is broadly reproduced in terms of not only the slope but also the absolute value.
The difference between the results with $\fdense=0.5$ and 0.1 is small,  
because the dust formation is dominated by stellar production at this early age. 

For the galaxies in Category II compared with the simulation result at $t=1$ Gyr
in Fig.~\ref{fig:dust_tot}b, the slope of the radial profile is reproduced well, 
although the simulated $\Dtot$ tend to be lower than the observed data points. 
However, the dust-to-metal ratio is reproduced well over the entire
radii (see Section~\ref{sec:depletion});  
thus, we suspect that the metallicity evolution implemented in our
model underproduces the metallicities for those galaxies. 
It requires a further check in the future, but this could simply be regarded as an  
early transient feature in isolated galaxy simulations of this type that does not include any 
cosmological evolution. 

The galaxies in Category III are compared with the simulation results at $t=3.0$ Gyr
in Fig.~\ref{fig:dust_tot}c. 
Although the observed data points have large dispersion,
the simulations roughly reproduce both the absolute value and slope of 
except the outer part of NGC 2403.
Even when we adopt the smaller $\fdense$, 
the result changes only by 30 \% and the difference is 
smaller than most observational uncertainties.

The simulation results at $t=10$ Gyr are compared with the galaxies in Category IV
in Fig.~\ref{fig:dust_tot}d.
The simulated slope is shallower at $t=10$ Gyr than 
at 3.0 Gyr, because the increase of $\Dtot$ by accretion is 
efficient up to the outer radii. 
The simulation result reproduces the overall trend of observed data
except for the central part of NGC 3031. 
Since the deficiency of dust abundance in NGC 3031 is
also seen in the dust-to-metal ratio (see Section~\ref{sec:depletion}),
the dust destruction in the central part of 
NGC 3031 may be much more efficient than assumed in this paper.
Because the accretion is saturated at this late age, 
the dependence of simulation result on $\fdense$ is very weak.


\subsection{Depletion}
\label{sec:depletion}

\begin{figure}
\includegraphics[width=7.7cm]{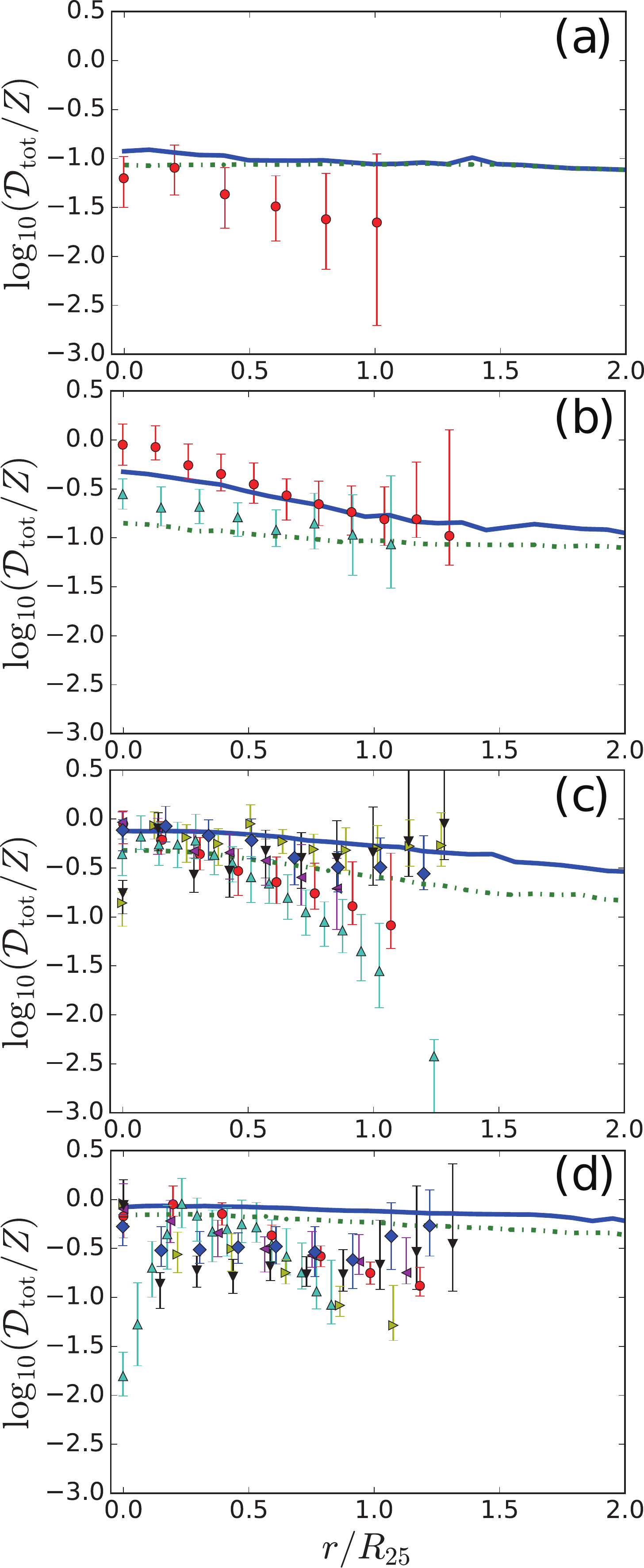}
\caption{
Comparison of simulation results with 
observed radial profiles of depletion in each category [panels ({\it a}) -- ({\it d}) 
correspond to Category I to IV, respectively]. 
The solid and dotted lines represent
the cases with $f_{\rm dense} = 0.5 $ and $0.1$, respectively.
The data points are obtained by \citet{2012MNRAS.423...38M}.
The symbols for observational points indicate the same galaxies 
as Fig. \ \ref{fig:dust_tot}.
}
\label{fig:depletion}
\end{figure}

We show the radial profiles of depletion (i.e., dust-to-metal ratio), 
which is defined as $\Dtot / Z$. 
We compare the simulated radial profile of $\Dtot /Z$ in Fig.~\ref{fig:depletion}, 
similarly to Fig.~\ref{fig:dust_tot}.
We calculate the depletion at each radius by
\begin{eqnarray} 
\dfrac{\mathcal{D}_{\rm tot}}{Z}(r)\equiv 
\dfrac{\displaystyle\sum_{i}m_{i}^{\rm gas}\mathcal{D}_{{\rm tot},i}}
{\displaystyle\sum_{i} m_{i}^{\rm gas}Z_{i}^{\rm gas}}~,
\end{eqnarray}
where the two mass-weighted values $m_{i}^{\rm gas} \mathcal{D}_{{\rm tot},i}$ and
$m_{i}^{\rm gas} Z_{i}^{\rm gas}$ are summed up in each ring
for all SPH particles in the appropriate radial range. 
We use 100 bins on the linear scale in $0 \leq r/R_{25} \leq 2 $.

We find that the depletion is almost constant in the early phase at $t=0.3$ Gyr
and the models broadly reproduce the observed depletion in Holmberg II (Category I).
At this stage, the level of depletion is broadly determined 
by the condensation efficiency in stellar ejecta. 
Thus the depletion is almost constant ($\simeq f_{\rm in} = 0.1$).
Since accretion is not efficient at this stage,
the simulation result is almost independent of $f_{\rm dense}$. 

For the galaxies in Category II that are compared with the simulation at
$t=1$ Gyr in Fig.~\ref{fig:depletion}b, 
we find that the depletion profile is broadly consistent with the
observational data. As interpreted by \cite{2012MNRAS.423...38M},
the decreasing trend of depletion with increasing radius 
is caused by more efficient accretion in the centre than in the outer regions, 
since dust growth by accretion is more efficient in more metal-rich environment.
When we adopt $\fdense =0.1$, we underestimate the depletion in the central region. 

For the galaxies in Category III compared with the simulation at $t=3$ Gyr
in Fig.~\ref{fig:depletion}c, 
the simulation somewhat overpredicts depletion against observed data. 
The radial gradient of depletion is shallower in Category III than in Category II, 
because the effect of accretion becomes more prevalent up to larger radii at later times.
When we consider the case of $\fdense =0.1$, 
the discrepancy between the simulation and observed data is less. 
NGC 2403 shows a rapid drop of depletion at the outer radii, 
which cannot be explained by our models, however 
it is still consistent with our model up to $r/R_{25}\sim 0.5$ .

Finally, we compare the galaxies in Category IV with the simulation at $t=10$ Gyr
in Fig.~\ref{fig:depletion}d.
Our simulation broadly overestimates the observed data, 
and the slope is flatter than the observed data. 
The radial profile of depletion and $\Dtot$ is flatter at $t=10$ Gyr than at 3 Gyr. 
Because accretion is saturated at later times, 
the difference of $\fdense$ does not affect the results significantly.
Some galaxies in Category IV indeed have a flat profile of $\Dtot/Z$,
which is qualitatively consistent with our results.

Overall, our simulated galaxy exhibits increasing depletion as a function of time
from $\Dtot / Z \sim -1.0$ to $0.0$\,dex.  We note that \citet[][]{2016arXiv160700288W} and  
\citet[][]{2016arXiv160808621D} reported a similar range of depletion for 
both QSO-DLAs and GRB-DLAs, and that depletion is a strong function of metallicity. 
These observational results support our conclusion that grain growth in the ISM is important
as we already saw in Fig.\,\ref{fig_dust-gasMassRatio}.


\section{Conclusion}
In this paper, we investigate the time evolution and spatial distribution of dust in an isolated galaxy with a modified version of \texttt{GADGET-3} SPH code. 
To represent the grain size distribution, we calculate the abundances 
of small and large grains separately based on the one-zone model 
by \citet[][]{2015MNRAS.447.2937H}.
Dust production by stars and destruction by SNe are implemented
consistently with star formation.
In order to overcome the mass resolution of the simulation,
we develop subgrid models for accretion and coagulation, since
these processes occur in dense clouds whose sizes are
below the spatial resolution of the simulation.

We find that the assumption of fixed dust-to-metal mass ratio becomes 
no longer valid when the galaxy age is $\gtrsim 0.2$ Gyr,  because
the grain growth by accretion produces a nonlinear dependence of
dust-to-gas ratio on the metallicity.  Small grain production by shattering
triggers accretion, because small grains are more efficient in accreting the
gas-phase metals.
In addition, coagulation becomes also significant at ages $\ga $1 Gyr after
a large amount of small grains are produced by accretion.
The age-dependent contributions of those processes are all important
in driving the evolution of dust-to-gas ratio and grain size distribution
at various epochs, and thus should be included in any calculation of
dust evolution in galaxies.

Finally, we made a first attempt of comparing our simulation results with
spatially resolved observational data of nearby galaxies. 
To extract the typical age in a simple way, we use the sSFR as an indicator, 
and compared the observed radial profiles of dust-to-gas ratio
and dust-to-metal ratio (i.e., depletion) with the simulation snapshots at various ages. 
We find that the radial profiles calculated in our models are broadly consistent with
observations. The negative radial gradient of dust-to-gas ratio is explained by
the tight relation between dust-to-gas ratio and metallicity.
The radial profile of depletion is flat at early epochs ($t \lesssim 0.3$ Gyr),  
because the dust-to-metal ratio is simply determined 
by the dust condensation efficiency in stellar ejecta.
At later stages, the radial gradient of depletion is negative, 
which represents the fact that the
dust-to-gas ratio is a nonlinear (strong) function of metallicity due to
accretion. We also reproduce the observational trend that the radial gradient
of depletion becomes flatter as the galaxy is aged, 
which indicates that the regions with efficient grain growth
by accretion extend from inside to outwards.

\section*{Acknowledgement}
We are grateful to Volker Springel for providing us with the original version of GADGET-3 code,
and to Akio Inoue and Hidenobu Yajima for useful discussions.
This work was supported in part by JSPS KAKENHI Grant Number 26247022.
HH thanks the Ministry of Science and Technology for support through grant \
MOST 105-2112-M-001-027-MY3. 
Numerical simulations were in part carried out on XC30 at the Centre for Computational Astrophysics, National Astronomical Observatory of Japan. 



\bibliographystyle{mnras}
\bibliography{ken} 

\begin{thebibliography}{}
\makeatletter
\relax
\def\mn@urlcharsother{\let\do\@makeother \do\$\do\&\do\#\do\^\do\_\do\%\do\~}
\def\mn@doi{\begingroup\mn@urlcharsother \@ifnextchar [ {\mn@doi@}
  {\mn@doi@[]}}
\def\mn@doi@[#1]#2{\def\@tempa{#1}\ifx\@tempa\@empty \href
  {http://dx.doi.org/#2} {doi:#2}\else \href {http://dx.doi.org/#2} {#1}\fi
  \endgroup}
\def\mn@eprint#1#2{\mn@eprint@#1:#2::\@nil}
\def\mn@eprint@arXiv#1{\href {http://arxiv.org/abs/#1} {{\tt arXiv:#1}}}
\def\mn@eprint@dblp#1{\href {http://dblp.uni-trier.de/rec/bibtex/#1.xml}
  {dblp:#1}}
\def\mn@eprint@#1:#2:#3:#4\@nil{\def\@tempa {#1}\def\@tempb {#2}\def\@tempc
  {#3}\ifx \@tempc \@empty \let \@tempc \@tempb \let \@tempb \@tempa \fi \ifx
  \@tempb \@empty \def\@tempb {arXiv}\fi \@ifundefined
  {mn@eprint@\@tempb}{\@tempb:\@tempc}{\expandafter \expandafter \csname
  mn@eprint@\@tempb\endcsname \expandafter{\@tempc}}}

\bibitem[\protect\citeauthoryear{{Agertz}, {Kravtsov}, {Leitner}  \&
  {Gnedin}}{{Agertz} et~al.}{2013}]{2013ApJ...770...25A}
{Agertz} O.,  {Kravtsov} A.~V.,  {Leitner} S.~N.,   {Gnedin} N.~Y.,  2013,
  \mn@doi [\apj] {10.1088/0004-637X/770/1/25}, \href
  {http://adsabs.harvard.edu/abs/2013ApJ...770...25A} {770, 25}

\bibitem[\protect\citeauthoryear{{Asano}, {Takeuchi}, {Hirashita}  \&
  {Inoue}}{{Asano} et~al.}{2013a}]{2013EP&S...65..213A}
{Asano} R.~S.,  {Takeuchi} T.~T.,  {Hirashita} H.,   {Inoue} A.~K.,  2013a,
  \mn@doi [Earth, Planets, and Space] {10.5047/eps.2012.04.014}, \href
  {http://adsabs.harvard.edu/abs/2013EP%26S...65..213A} {65, 213}

\bibitem[\protect\citeauthoryear{{Asano}, {Takeuchi}, {Hirashita}  \&
  {Nozawa}}{{Asano} et~al.}{2013b}]{2013MNRAS.432..637A}
{Asano} R.~S.,  {Takeuchi} T.~T.,  {Hirashita} H.,   {Nozawa} T.,  2013b,
  \mn@doi [\mnras] {10.1093/mnras/stt506}, \href
  {http://adsabs.harvard.edu/abs/2013MNRAS.432..637A} {432, 637}

\bibitem[\protect\citeauthoryear{{Asano}, {Takeuchi}, {Hirashita}  \&
  {Nozawa}}{{Asano} et~al.}{2014}]{2014MNRAS.440..134A}
{Asano} R.~S.,  {Takeuchi} T.~T.,  {Hirashita} H.,   {Nozawa} T.,  2014,
  \mn@doi [\mnras] {10.1093/mnras/stu208}, \href
  {http://adsabs.harvard.edu/abs/2014MNRAS.440..134A} {440, 134}

\bibitem[\protect\citeauthoryear{{Barlow} \& {Silk}}{{Barlow} \&
  {Silk}}{1976}]{1976ApJ...207..131B}
{Barlow} M.~J.,  {Silk} J.,  1976, \mn@doi [\apj] {10.1086/154477}, \href
  {http://adsabs.harvard.edu/abs/1976ApJ...207..131B} {207, 131}

\bibitem[\protect\citeauthoryear{{Bekki}}{{Bekki}}{2013}]{2013MNRAS.432.2298B}
{Bekki} K.,  2013, \mn@doi [\mnras] {10.1093/mnras/stt589}, \href
  {http://adsabs.harvard.edu/abs/2013MNRAS.432.2298B} {432, 2298}

\bibitem[\protect\citeauthoryear{{Bekki}}{{Bekki}}{2015}]{2015MNRAS.449.1625B}
{Bekki} K.,  2015, \mn@doi [\mnras] {10.1093/mnras/stv165}, \href
  {http://adsabs.harvard.edu/abs/2015MNRAS.449.1625B} {449, 1625}

\bibitem[\protect\citeauthoryear{{Bohren}, {Huffman}  \& {Kam}}{{Bohren}
  et~al.}{1983}]{1983Natur.306..625B}
{Bohren} C.~F.,  {Huffman} D.~R.,   {Kam} Z.,  1983, \nat, \href
  {http://adsabs.harvard.edu/abs/1983Natur.306..625B} {306, 625}

\bibitem[\protect\citeauthoryear{{Bryan} et~al.,}{{Bryan}
  et~al.}{2014}]{2014ApJS..211...19B}
{Bryan} G.~L.,  et~al., 2014, \mn@doi [\apjs] {10.1088/0067-0049/211/2/19},
  \href {http://adsabs.harvard.edu/abs/2014ApJS..211...19B} {211, 19}

\bibitem[\protect\citeauthoryear{{Buat}, {Boselli}, {Gavazzi}  \&
  {Bonfanti}}{{Buat} et~al.}{2002}]{2002A&A...383..801B}
{Buat} V.,  {Boselli} A.,  {Gavazzi} G.,   {Bonfanti} C.,  2002, \mn@doi [\aap]
  {10.1051/0004-6361:20011832}, \href
  {http://adsabs.harvard.edu/abs/2002A%26A...383..801B} {383, 801}

\bibitem[\protect\citeauthoryear{{Calzetti}, {Armus}, {Bohlin}, {Kinney},
  {Koornneef}  \& {Storchi-Bergmann}}{{Calzetti}
  et~al.}{2000}]{2000ApJ...533..682C}
{Calzetti} D.,  {Armus} L.,  {Bohlin} R.~C.,  {Kinney} A.~L.,  {Koornneef} J.,
   {Storchi-Bergmann} T.,  2000, \mn@doi [\apj] {10.1086/308692}, \href
  {http://adsabs.harvard.edu/abs/2000ApJ...533..682C} {533, 682}

\bibitem[\protect\citeauthoryear{{Cazaux} \& {Spaans}}{{Cazaux} \&
  {Spaans}}{2009}]{2009A&A...496..365C}
{Cazaux} S.,  {Spaans} M.,  2009, \mn@doi [\aap] {10.1051/0004-6361:200811302},
  \href {http://ads.nao.ac.jp/abs/2009A%26A...496..365C} {496, 365}

\bibitem[\protect\citeauthoryear{{Cazaux} \& {Tielens}}{{Cazaux} \&
  {Tielens}}{2004}]{2004ApJ...604..222C}
{Cazaux} S.,  {Tielens} A.~G.~G.~M.,  2004, \mn@doi [\apj] {10.1086/381775},
  \href {http://ads.nao.ac.jp/abs/2004ApJ...604..222C} {604, 222}

\bibitem[\protect\citeauthoryear{{Chabrier}}{{Chabrier}}{2003}]{2003PASP..115..763C}
{Chabrier} G.,  2003, \mn@doi [\pasp] {10.1086/376392}, \href
  {http://adsabs.harvard.edu/abs/2003PASP..115..763C} {115, 763}

\bibitem[\protect\citeauthoryear{{Chevalier}}{{Chevalier}}{1974}]{1974ApJ...188..501C}
{Chevalier} R.~A.,  1974, \mn@doi [\apj] {10.1086/152740}, \href
  {http://adsabs.harvard.edu/abs/1974ApJ...188..501C} {188, 501}

\bibitem[\protect\citeauthoryear{{Choi} \& {Nagamine}}{{Choi} \&
  {Nagamine}}{2009}]{2009MNRAS.393.1595C}
{Choi} J.-H.,  {Nagamine} K.,  2009, \mn@doi [\mnras]
  {10.1111/j.1365-2966.2008.14297.x}, \href
  {http://adsabs.harvard.edu/abs/2009MNRAS.393.1595C} {393, 1595}

\bibitem[\protect\citeauthoryear{{Choi} \& {Nagamine}}{{Choi} \&
  {Nagamine}}{2012}]{2012MNRAS.419.1280C}
{Choi} J.-H.,  {Nagamine} K.,  2012, \mn@doi [\mnras]
  {10.1111/j.1365-2966.2011.19788.x}, \href
  {http://adsabs.harvard.edu/abs/2012MNRAS.419.1280C} {419, 1280}

\bibitem[\protect\citeauthoryear{{De Cia}, {Ledoux}, {Mattsson}, {Petitjean},
  {Srianand}, {Gavignaud}  \& {Jenkins}}{{De Cia}
  et~al.}{2016}]{2016arXiv160808621D}
{De Cia} A.,  {Ledoux} C.,  {Mattsson} L.,  {Petitjean} P.,  {Srianand} R.,
  {Gavignaud} I.,   {Jenkins} E.~B.,  2016, preprint, \href
  {http://adsabs.harvard.edu/abs/2016arXiv160808621D} {} (\mn@eprint {arXiv}
  {1608.08621})

\bibitem[\protect\citeauthoryear{{Draine} \& {Li}}{{Draine} \&
  {Li}}{2001}]{2001ApJ...551..807D}
{Draine} B.~T.,  {Li} A.,  2001, \mn@doi [\apj] {10.1086/320227}, \href
  {http://adsabs.harvard.edu/abs/2001ApJ...551..807D} {551, 807}

\bibitem[\protect\citeauthoryear{{Durier} \& {Dalla Vecchia}}{{Durier} \&
  {Dalla Vecchia}}{2012}]{2012MNRAS.419..465D}
{Durier} F.,  {Dalla Vecchia} C.,  2012, \mn@doi [\mnras]
  {10.1111/j.1365-2966.2011.19712.x}, \href
  {http://adsabs.harvard.edu/abs/2012MNRAS.419..465D} {419, 465}

\bibitem[\protect\citeauthoryear{{Dutta}, {Begum}, {Bharadwaj}  \&
  {Chengalur}}{{Dutta} et~al.}{2013}]{2013NewA...19...89D}
{Dutta} P.,  {Begum} A.,  {Bharadwaj} S.,   {Chengalur} J.~N.,  2013, \mn@doi
  [\na] {10.1016/j.newast.2012.08.008}, \href
  {http://adsabs.harvard.edu/abs/2013NewA...19...89D} {19, 89}

\bibitem[\protect\citeauthoryear{{Dwek}}{{Dwek}}{1998}]{1998ApJ...501..643D}
{Dwek} E.,  1998, \mn@doi [\apj] {10.1086/305829}, \href
  {http://adsabs.harvard.edu/abs/1998ApJ...501..643D} {501, 643}

\bibitem[\protect\citeauthoryear{{Elmegreen}}{{Elmegreen}}{1998}]{1998ggs..book.....E}
{Elmegreen} D.~M.,  1998, {Galaxies and galactic structure}.
{Upper Saddle River, NJ : Prentice Hall}

\bibitem[\protect\citeauthoryear{{Fall}, {Krumholz}  \& {Matzner}}{{Fall}
  et~al.}{2010}]{2010ApJ...710L.142F}
{Fall} S.~M.,  {Krumholz} M.~R.,   {Matzner} C.~D.,  2010, \mn@doi [\apjl]
  {10.1088/2041-8205/710/2/L142}, \href
  {http://adsabs.harvard.edu/abs/2010ApJ...710L.142F} {710, L142}

\bibitem[\protect\citeauthoryear{{Gould} \& {Salpeter}}{{Gould} \&
  {Salpeter}}{1963}]{1963ApJ...138..393G}
{Gould} R.~J.,  {Salpeter} E.~E.,  1963, \mn@doi [\apj] {10.1086/147654}, \href
  {http://adsabs.harvard.edu/abs/1963ApJ...138..393G} {138, 393}

\bibitem[\protect\citeauthoryear{{Heesen}, {Brinks}, {Leroy}, {Heald}, {Braun},
  {Bigiel}  \& {Beck}}{{Heesen} et~al.}{2014}]{2014AJ....147..103H}
{Heesen} V.,  {Brinks} E.,  {Leroy} A.~K.,  {Heald} G.,  {Braun} R.,  {Bigiel}
  F.,   {Beck} R.,  2014, \mn@doi [\aj] {10.1088/0004-6256/147/5/103}, \href
  {http://adsabs.harvard.edu/abs/2014AJ....147..103H} {147, 103}

\bibitem[\protect\citeauthoryear{{Hirashita}}{{Hirashita}}{2015}]{2015MNRAS.447.2937H}
{Hirashita} H.,  2015, \mn@doi [\mnras] {10.1093/mnras/stu2617}, \href
  {http://adsabs.harvard.edu/abs/2015MNRAS.447.2937H} {447, 2937}

\bibitem[\protect\citeauthoryear{{Hirashita} \& {Kuo}}{{Hirashita} \&
  {Kuo}}{2011}]{2011MNRAS.416.1340H}
{Hirashita} H.,  {Kuo} T.-M.,  2011, \mn@doi [\mnras]
  {10.1111/j.1365-2966.2011.19131.x}, \href
  {http://adsabs.harvard.edu/abs/2011MNRAS.416.1340H} {416, 1340}

\bibitem[\protect\citeauthoryear{{Hirashita} \& {Voshchinnikov}}{{Hirashita} \&
  {Voshchinnikov}}{2014}]{2014MNRAS.437.1636H}
{Hirashita} H.,  {Voshchinnikov} N.~V.,  2014, \mn@doi [\mnras]
  {10.1093/mnras/stt1997}, \href
  {http://adsabs.harvard.edu/abs/2014MNRAS.437.1636H} {437, 1636}

\bibitem[\protect\citeauthoryear{{Hirashita} \& {Yan}}{{Hirashita} \&
  {Yan}}{2009}]{2009MNRAS.394.1061H}
{Hirashita} H.,  {Yan} H.,  2009, \mn@doi [\mnras]
  {10.1111/j.1365-2966.2009.14405.x}, \href
  {http://adsabs.harvard.edu/abs/2009MNRAS.394.1061H} {394, 1061}

\bibitem[\protect\citeauthoryear{{Hopkins}}{{Hopkins}}{2013}]{2013MNRAS.428.2840H}
{Hopkins} P.~F.,  2013, \mn@doi [\mnras] {10.1093/mnras/sts210}, \href
  {http://adsabs.harvard.edu/abs/2013MNRAS.428.2840H} {428, 2840}

\bibitem[\protect\citeauthoryear{{Hopkins}, {Quataert}  \& {Murray}}{{Hopkins}
  et~al.}{2011}]{2011MNRAS.417..950H}
{Hopkins} P.~F.,  {Quataert} E.,   {Murray} N.,  2011, \mn@doi [\mnras]
  {10.1111/j.1365-2966.2011.19306.x}, \href
  {http://adsabs.harvard.edu/abs/2011MNRAS.417..950H} {417, 950}

\bibitem[\protect\citeauthoryear{{Hou}, {Hirashita}  \& {Micha{\l}owski}}{{Hou}
  et~al.}{2016}]{2016arXiv160806099H}
{Hou} K.-C.,  {Hirashita} H.,   {Micha{\l}owski} M.~J.,  2016, preprint, \href
  {http://adsabs.harvard.edu/abs/2016arXiv160806099H} {} (\mn@eprint {arXiv}
  {1608.06099})

\bibitem[\protect\citeauthoryear{{Inoue}}{{Inoue}}{2003}]{2003PASJ...55..901I}
{Inoue} A.~K.,  2003, \mn@doi [\pasj] {10.1093/pasj/55.5.901}, \href
  {http://adsabs.harvard.edu/abs/2003PASJ...55..901I} {55, 901}

\bibitem[\protect\citeauthoryear{{Inoue}}{{Inoue}}{2011a}]{2011EP&S...63.1027I}
{Inoue} A.~K.,  2011a, \mn@doi [Earth, Planets, and Space]
  {10.5047/eps.2011.02.013}, \href
  {http://adsabs.harvard.edu/abs/2011EP%26S...63.1027I} {63, 1027}

\bibitem[\protect\citeauthoryear{{Inoue}}{{Inoue}}{2011b}]{2011MNRAS.415.2920I}
{Inoue} A.~K.,  2011b, \mn@doi [\mnras] {10.1111/j.1365-2966.2011.18906.x},
  \href {http://adsabs.harvard.edu/abs/2011MNRAS.415.2920I} {415, 2920}

\bibitem[\protect\citeauthoryear{{Jaacks}, {Nagamine}  \& {Choi}}{{Jaacks}
  et~al.}{2012}]{2012MNRAS.427..403J}
{Jaacks} J.,  {Nagamine} K.,   {Choi} J.~H.,  2012, \mn@doi [\mnras]
  {10.1111/j.1365-2966.2012.21989.x}, \href
  {http://adsabs.harvard.edu/abs/2012MNRAS.427..403J} {427, 403}

\bibitem[\protect\citeauthoryear{{Jaacks}, {Thompson}  \& {Nagamine}}{{Jaacks}
  et~al.}{2013}]{2013ApJ...766...94J}
{Jaacks} J.,  {Thompson} R.,   {Nagamine} K.,  2013, \mn@doi [\apj]
  {10.1088/0004-637X/766/2/94}, \href
  {http://adsabs.harvard.edu/abs/2013ApJ...766...94J} {766, 94}

\bibitem[\protect\citeauthoryear{{Jenkins}}{{Jenkins}}{2009}]{2009ApJ...700.1299J}
{Jenkins} E.~B.,  2009, \mn@doi [\apj] {10.1088/0004-637X/700/2/1299}, \href
  {http://adsabs.harvard.edu/abs/2009ApJ...700.1299J} {700, 1299}

\bibitem[\protect\citeauthoryear{{Jones}, {Tielens}, {Hollenbach}  \&
  {McKee}}{{Jones} et~al.}{1994}]{1994ApJ...433..797J}
{Jones} A.~P.,  {Tielens} A.~G.~G.~M.,  {Hollenbach} D.~J.,   {McKee} C.~F.,
  1994, \mn@doi [\apj] {10.1086/174689}, \href
  {http://adsabs.harvard.edu/abs/1994ApJ...433..797J} {433, 797}

\bibitem[\protect\citeauthoryear{{Jones}, {Tielens}  \& {Hollenbach}}{{Jones}
  et~al.}{1996}]{1996ApJ...469..740J}
{Jones} A.~P.,  {Tielens} A.~G.~G.~M.,   {Hollenbach} D.~J.,  1996, \mn@doi
  [\apj] {10.1086/177823}, \href
  {http://adsabs.harvard.edu/abs/1996ApJ...469..740J} {469, 740}

\bibitem[\protect\citeauthoryear{{Kataoka}, {Okuzumi}, {Tanaka}  \&
  {Nomura}}{{Kataoka} et~al.}{2014}]{2014A&A...568A..42K}
{Kataoka} A.,  {Okuzumi} S.,  {Tanaka} H.,   {Nomura} H.,  2014, \mn@doi [\aap]
  {10.1051/0004-6361/201323199}, \href
  {http://adsabs.harvard.edu/abs/2014A\%26A...568A..42K} {568, A42}

\bibitem[\protect\citeauthoryear{{Kennicutt} \& {Evans}}{{Kennicutt} \&
  {Evans}}{2012}]{2012ARA&A..50..531K}
{Kennicutt} R.~C.,  {Evans} N.~J.,  2012, \mn@doi [\araa]
  {10.1146/annurev-astro-081811-125610}, \href
  {http://adsabs.harvard.edu/abs/2012ARA%26A..50..531K} {50, 531}

\bibitem[\protect\citeauthoryear{{Kim} et~al.,}{{Kim}
  et~al.}{2014}]{2014ApJS..210...14K}
{Kim} J.-h.,  et~al., 2014, \mn@doi [\apjs] {10.1088/0067-0049/210/1/14}, \href
  {http://adsabs.harvard.edu/abs/2014ApJS..210...14K} {210, 14}

\bibitem[\protect\citeauthoryear{{Kuo} \& {Hirashita}}{{Kuo} \&
  {Hirashita}}{2012}]{2012MNRAS.424L..34K}
{Kuo} T.-M.,  {Hirashita} H.,  2012, \mn@doi [\mnras]
  {10.1111/j.1745-3933.2012.01282.x}, \href
  {http://adsabs.harvard.edu/abs/2012MNRAS.424L..34K} {424, L34}

\bibitem[\protect\citeauthoryear{{Kuo}, {Hirashita}  \& {Zafar}}{{Kuo}
  et~al.}{2013}]{2013MNRAS.436.1238K}
{Kuo} T.-M.,  {Hirashita} H.,   {Zafar} T.,  2013, \mn@doi [\mnras]
  {10.1093/mnras/stt1648}, \href
  {http://adsabs.harvard.edu/abs/2013MNRAS.436.1238K} {436, 1238}

\bibitem[\protect\citeauthoryear{{Larson}}{{Larson}}{2005}]{2005MNRAS.359..211L}
{Larson} R.~B.,  2005, \mn@doi [\mnras] {10.1111/j.1365-2966.2005.08881.x},
  \href {http://adsabs.harvard.edu/abs/2005MNRAS.359..211L} {359, 211}

\bibitem[\protect\citeauthoryear{{Li} \& {Draine}}{{Li} \&
  {Draine}}{2001}]{2001ApJ...554..778L}
{Li} A.,  {Draine} B.~T.,  2001, \mn@doi [\apj] {10.1086/323147}, \href
  {http://adsabs.harvard.edu/abs/2001ApJ...554..778L} {554, 778}

\bibitem[\protect\citeauthoryear{{Mathis}}{{Mathis}}{1990}]{1990ARA&A..28...37M}
{Mathis} J.~S.,  1990, \mn@doi [\araa] {10.1146/annurev.aa.28.090190.000345},
  \href {http://adsabs.harvard.edu/abs/1990ARA%26A..28...37M} {28, 37}

\bibitem[\protect\citeauthoryear{{Mathis}, {Rumpl}  \& {Nordsieck}}{{Mathis}
  et~al.}{1977}]{1977ApJ...217..425M}
{Mathis} J.~S.,  {Rumpl} W.,   {Nordsieck} K.~H.,  1977, \mn@doi [\apj]
  {10.1086/155591}, \href {http://adsabs.harvard.edu/abs/1977ApJ...217..425M}
  {217, 425}

\bibitem[\protect\citeauthoryear{{Mattsson} \& {Andersen}}{{Mattsson} \&
  {Andersen}}{2012}]{2012MNRAS.423...38M}
{Mattsson} L.,  {Andersen} A.~C.,  2012, \mn@doi [\mnras]
  {10.1111/j.1365-2966.2012.20574.x}, \href
  {http://adsabs.harvard.edu/abs/2012MNRAS.423...38M} {423, 38}

\bibitem[\protect\citeauthoryear{{McKee}}{{McKee}}{1989}]{1989IAUS..135..431M}
{McKee} C.,  1989, in {Allamandola} L.~J.,  {Tielens} A.~G.~G.~M.,  eds,  IAU
  Symposium Vol. 135, Interstellar Dust. p.~431

\bibitem[\protect\citeauthoryear{{McKee} \& {Ostriker}}{{McKee} \&
  {Ostriker}}{1977}]{1977ApJ...218..148M}
{McKee} C.~F.,  {Ostriker} J.~P.,  1977, \mn@doi [\apj] {10.1086/155667}, \href
  {http://adsabs.harvard.edu/abs/1977ApJ...218..148M} {218, 148}

\bibitem[\protect\citeauthoryear{{McKee}, {Hollenbach}, {Seab}  \&
  {Tielens}}{{McKee} et~al.}{1987}]{1987ApJ...318..674M}
{McKee} C.~F.,  {Hollenbach} D.~J.,  {Seab} G.~C.,   {Tielens} A.~G.~G.~M.,
  1987, \mn@doi [\apj] {10.1086/165403}, \href
  {http://adsabs.harvard.edu/abs/1987ApJ...318..674M} {318, 674}

\bibitem[\protect\citeauthoryear{{McKinnon}, {Torrey}, {Vogelsberger},
  {Hayward}  \& {Marinacci}}{{McKinnon} et~al.}{2016a}]{2016arXiv160602714M}
{McKinnon} R.,  {Torrey} P.,  {Vogelsberger} M.,  {Hayward} C.~C.,
  {Marinacci} F.,  2016a, preprint, \href
  {http://adsabs.harvard.edu/abs/2016arXiv160602714M} {} (\mn@eprint {arXiv}
  {1606.02714})

\bibitem[\protect\citeauthoryear{{McKinnon}, {Torrey}  \&
  {Vogelsberger}}{{McKinnon} et~al.}{2016b}]{2016MNRAS.457.3775M}
{McKinnon} R.,  {Torrey} P.,   {Vogelsberger} M.,  2016b, \mn@doi [\mnras]
  {10.1093/mnras/stw253}, \href
  {http://adsabs.harvard.edu/abs/2016MNRAS.457.3775M} {457, 3775}

\bibitem[\protect\citeauthoryear{{Morris}}{{Morris}}{1996}]{1996PASA...13...97M}
{Morris} J.~P.,  1996, \pasa, \href
  {http://adsabs.harvard.edu/abs/1996PASA...13...97M} {13, 97}

\bibitem[\protect\citeauthoryear{{Moustakas}, {Kennicutt}, {Tremonti}, {Dale},
  {Smith}  \& {Calzetti}}{{Moustakas} et~al.}{2010}]{2010ApJS..190..233M}
{Moustakas} J.,  {Kennicutt} Jr. R.~C.,  {Tremonti} C.~A.,  {Dale} D.~A.,
  {Smith} J.-D.~T.,   {Calzetti} D.,  2010, \mn@doi [\apjs]
  {10.1088/0067-0049/190/2/233}, \href
  {http://adsabs.harvard.edu/abs/2010ApJS..190..233M} {190, 233}

\bibitem[\protect\citeauthoryear{{Nagamine}, {Fukugita}, {Cen}  \&
  {Ostriker}}{{Nagamine} et~al.}{2001}]{2001ApJ...558..497N}
{Nagamine} K.,  {Fukugita} M.,  {Cen} R.,   {Ostriker} J.~P.,  2001, \mn@doi
  [\apj] {10.1086/322293}, \href
  {http://adsabs.harvard.edu/abs/2001ApJ...558..497N} {558, 497}

\bibitem[\protect\citeauthoryear{{Nagamine}, {Springel}, {Hernquist}  \&
  {Machacek}}{{Nagamine} et~al.}{2004}]{2004MNRAS.350..385N}
{Nagamine} K.,  {Springel} V.,  {Hernquist} L.,   {Machacek} M.,  2004, \mn@doi
  [\mnras] {10.1111/j.1365-2966.2004.07664.x}, \href
  {http://adsabs.harvard.edu/abs/2004MNRAS.350..385N} {350, 385}

\bibitem[\protect\citeauthoryear{{Nozawa} \& {Fukugita}}{{Nozawa} \&
  {Fukugita}}{2013}]{2013ApJ...770...27N}
{Nozawa} T.,  {Fukugita} M.,  2013, \mn@doi [\apj]
  {10.1088/0004-637X/770/1/27}, \href
  {http://adsabs.harvard.edu/abs/2013ApJ...770...27N} {770, 27}

\bibitem[\protect\citeauthoryear{{Nozawa}, {Kozasa}  \& {Habe}}{{Nozawa}
  et~al.}{2006}]{2006ApJ...648..435N}
{Nozawa} T.,  {Kozasa} T.,   {Habe} A.,  2006, \mn@doi [\apj] {10.1086/505639},
  \href {http://adsabs.harvard.edu/abs/2006ApJ...648..435N} {648, 435}

\bibitem[\protect\citeauthoryear{{Nozawa}, {Asano}, {Hirashita}  \&
  {Takeuchi}}{{Nozawa} et~al.}{2015}]{2015MNRAS.447L..16N}
{Nozawa} T.,  {Asano} R.~S.,  {Hirashita} H.,   {Takeuchi} T.~T.,  2015,
  \mn@doi [\mnras] {10.1093/mnrasl/slu175}, \href
  {http://adsabs.harvard.edu/abs/2015MNRAS.447L..16N} {447, L16}

\bibitem[\protect\citeauthoryear{{Okuzumi}, {Tanaka}  \& {Sakagami}}{{Okuzumi}
  et~al.}{2009}]{2009ApJ...707.1247O}
{Okuzumi} S.,  {Tanaka} H.,   {Sakagami} M.-a.,  2009, \mn@doi [\apj]
  {10.1088/0004-637X/707/2/1247}, \href
  {http://adsabs.harvard.edu/abs/2009ApJ...707.1247O} {707, 1247}

\bibitem[\protect\citeauthoryear{{Omukai}, {Tsuribe}, {Schneider}  \&
  {Ferrara}}{{Omukai} et~al.}{2005}]{2005ApJ...626..627O}
{Omukai} K.,  {Tsuribe} T.,  {Schneider} R.,   {Ferrara} A.,  2005, \mn@doi
  [\apj] {10.1086/429955}, \href
  {http://adsabs.harvard.edu/abs/2005ApJ...626..627O} {626, 627}

\bibitem[\protect\citeauthoryear{{Popping}, {Somerville}  \&
  {Galametz}}{{Popping} et~al.}{2016}]{2016arXiv160908622P}
{Popping} G.,  {Somerville} R.~S.,   {Galametz} M.,  2016, preprint, \href
  {http://adsabs.harvard.edu/abs/2016arXiv160908622P} {} (\mn@eprint {arXiv}
  {1609.08622})

\bibitem[\protect\citeauthoryear{{R{\'e}my-Ruyer} et~al.,}{{R{\'e}my-Ruyer}
  et~al.}{2014}]{2014A&A...563A..31R}
{R{\'e}my-Ruyer} A.,  et~al., 2014, \mn@doi [\aap]
  {10.1051/0004-6361/201322803}, \href
  {http://adsabs.harvard.edu/abs/2014A%26A...563A..31R} {563, A31}

\bibitem[\protect\citeauthoryear{{Saitoh} \& {Makino}}{{Saitoh} \&
  {Makino}}{2013}]{2013ApJ...768...44S}
{Saitoh} T.~R.,  {Makino} J.,  2013, \mn@doi [\apj]
  {10.1088/0004-637X/768/1/44}, \href
  {http://adsabs.harvard.edu/abs/2013ApJ...768...44S} {768, 44}

\bibitem[\protect\citeauthoryear{{Sandstrom} et~al.,}{{Sandstrom}
  et~al.}{2013}]{2013ApJ...777....5S}
{Sandstrom} K.~M.,  et~al., 2013, \mn@doi [\apj] {10.1088/0004-637X/777/1/5},
  \href {http://adsabs.harvard.edu/abs/2013ApJ...777....5S} {777, 5}

\bibitem[\protect\citeauthoryear{{Schaye} et~al.,}{{Schaye}
  et~al.}{2015}]{2015MNRAS.446..521S}
{Schaye} J.,  et~al., 2015, \mn@doi [\mnras] {10.1093/mnras/stu2058}, \href
  {http://adsabs.harvard.edu/abs/2015MNRAS.446..521S} {446, 521}

\bibitem[\protect\citeauthoryear{{Schneider}, {Omukai}, {Inoue}  \&
  {Ferrara}}{{Schneider} et~al.}{2006}]{2006MNRAS.369.1437S}
{Schneider} R.,  {Omukai} K.,  {Inoue} A.~K.,   {Ferrara} A.,  2006, \mn@doi
  [\mnras] {10.1111/j.1365-2966.2006.10391.x}, \href
  {http://adsabs.harvard.edu/abs/2006MNRAS.369.1437S} {369, 1437}

\bibitem[\protect\citeauthoryear{{Shimizu}, {Inoue}, {Okamoto}  \&
  {Yoshida}}{{Shimizu} et~al.}{2014}]{2014MNRAS.440..731S}
{Shimizu} I.,  {Inoue} A.~K.,  {Okamoto} T.,   {Yoshida} N.,  2014, \mn@doi
  [\mnras] {10.1093/mnras/stu265}, \href
  {http://adsabs.harvard.edu/abs/2014MNRAS.440..731S} {440, 731}

\bibitem[\protect\citeauthoryear{{Shimizu}, {Inoue}, {Yoshida}  \&
  {Okamoto}}{{Shimizu} et~al.}{2015}]{2015arXiv150900800S}
{Shimizu} I.,  {Inoue} A.~K.,  {Yoshida} N.,   {Okamoto} T.,  2015, preprint,
  \href {http://adsabs.harvard.edu/abs/2015arXiv150900800S} {} (\mn@eprint
  {arXiv} {1509.00800})

\bibitem[\protect\citeauthoryear{{Smith} et~al.,}{{Smith}
  et~al.}{2016}]{2016MNRAS.462..331S}
{Smith} M.~W.~L.,  et~al., 2016, \mn@doi [\mnras] {10.1093/mnras/stw1611},
  \href {http://adsabs.harvard.edu/abs/2016MNRAS.462..331S} {462, 331}

\bibitem[\protect\citeauthoryear{{Springel}}{{Springel}}{2005}]{2005MNRAS.364.1105S}
{Springel} V.,  2005, \mn@doi [\mnras] {10.1111/j.1365-2966.2005.09655.x},
  \href {http://adsabs.harvard.edu/abs/2005MNRAS.364.1105S} {364, 1105}

\bibitem[\protect\citeauthoryear{{Springel} \& {Hernquist}}{{Springel} \&
  {Hernquist}}{2002}]{2002MNRAS.333..649S}
{Springel} V.,  {Hernquist} L.,  2002, \mn@doi [\mnras]
  {10.1046/j.1365-8711.2002.05445.x}, \href
  {http://adsabs.harvard.edu/abs/2002MNRAS.333..649S} {333, 649}

\bibitem[\protect\citeauthoryear{{Springel} \& {Hernquist}}{{Springel} \&
  {Hernquist}}{2003}]{2003MNRAS.339..289S}
{Springel} V.,  {Hernquist} L.,  2003, \mn@doi [\mnras]
  {10.1046/j.1365-8711.2003.06206.x}, \href
  {http://adsabs.harvard.edu/abs/2003MNRAS.339..289S} {339, 289}

\bibitem[\protect\citeauthoryear{{Steidel}, {Adelberger}, {Giavalisco},
  {Dickinson}  \& {Pettini}}{{Steidel} et~al.}{1999}]{1999ApJ...519....1S}
{Steidel} C.~C.,  {Adelberger} K.~L.,  {Giavalisco} M.,  {Dickinson} M.,
  {Pettini} M.,  1999, \mn@doi [\apj] {10.1086/307363}, \href
  {http://adsabs.harvard.edu/abs/1999ApJ...519....1S} {519, 1}

\bibitem[\protect\citeauthoryear{{Stinson}, {Brook}, {Macci{\`o}}, {Wadsley},
  {Quinn}  \& {Couchman}}{{Stinson} et~al.}{2013}]{2013MNRAS.428..129S}
{Stinson} G.~S.,  {Brook} C.,  {Macci{\`o}} A.~V.,  {Wadsley} J.,  {Quinn}
  T.~R.,   {Couchman} H.~M.~P.,  2013, \mn@doi [\mnras] {10.1093/mnras/sts028},
  \href {http://adsabs.harvard.edu/abs/2013MNRAS.428..129S} {428, 129}

\bibitem[\protect\citeauthoryear{{Takeuchi}, {Buat}, {Heinis}, {Giovannoli},
  {Yuan}, {Iglesias-P{\'a}ramo}, {Murata}  \& {Burgarella}}{{Takeuchi}
  et~al.}{2010}]{2010A&A...514A...4T}
{Takeuchi} T.~T.,  {Buat} V.,  {Heinis} S.,  {Giovannoli} E.,  {Yuan} F.-T.,
  {Iglesias-P{\'a}ramo} J.,  {Murata} K.~L.,   {Burgarella} D.,  2010, \mn@doi
  [\aap] {10.1051/0004-6361/200913476}, \href
  {http://adsabs.harvard.edu/abs/2010A%26A...514A...4T} {514, A4}

\bibitem[\protect\citeauthoryear{{Takeuchi}, {Yuan}, {Ikeyama}, {Murata}  \&
  {Inoue}}{{Takeuchi} et~al.}{2012}]{2012ApJ...755..144T}
{Takeuchi} T.~T.,  {Yuan} F.-T.,  {Ikeyama} A.,  {Murata} K.~L.,   {Inoue}
  A.~K.,  2012, \mn@doi [\apj] {10.1088/0004-637X/755/2/144}, \href
  {http://adsabs.harvard.edu/abs/2012ApJ...755..144T} {755, 144}

\bibitem[\protect\citeauthoryear{{Thompson}, {Nagamine}, {Jaacks}  \&
  {Choi}}{{Thompson} et~al.}{2014}]{2014ApJ...780..145T}
{Thompson} R.,  {Nagamine} K.,  {Jaacks} J.,   {Choi} J.-H.,  2014, \mn@doi
  [\apj] {10.1088/0004-637X/780/2/145}, \href
  {http://adsabs.harvard.edu/abs/2014ApJ...780..145T} {780, 145}

\bibitem[\protect\citeauthoryear{{Todoroki}}{{Todoroki}}{2014}]{2014MsT..........1T}
{Todoroki} K.,  2014, Master's thesis, University of Nevada, Las Vegas

\bibitem[\protect\citeauthoryear{{Valiante}, {Schneider}, {Bianchi}  \&
  {Andersen}}{{Valiante} et~al.}{2009}]{2009MNRAS.397.1661V}
{Valiante} R.,  {Schneider} R.,  {Bianchi} S.,   {Andersen} A.~C.,  2009,
  \mn@doi [\mnras] {10.1111/j.1365-2966.2009.15076.x}, \href
  {http://adsabs.harvard.edu/abs/2009MNRAS.397.1661V} {397, 1661}

\bibitem[\protect\citeauthoryear{{Vogelsberger} et~al.,}{{Vogelsberger}
  et~al.}{2014}]{2014MNRAS.444.1518V}
{Vogelsberger} M.,  et~al., 2014, \mn@doi [\mnras] {10.1093/mnras/stu1536},
  \href {http://adsabs.harvard.edu/abs/2014MNRAS.444.1518V} {444, 1518}

\bibitem[\protect\citeauthoryear{{Voshchinnikov} \&
  {Hirashita}}{{Voshchinnikov} \& {Hirashita}}{2014}]{2014MNRAS.445..301V}
{Voshchinnikov} N.~V.,  {Hirashita} H.,  2014, \mn@doi [\mnras]
  {10.1093/mnras/stu1720}, \href
  {http://adsabs.harvard.edu/abs/2014MNRAS.445..301V} {445, 301}

\bibitem[\protect\citeauthoryear{{Whitworth}, {Boffin}  \&
  {Francis}}{{Whitworth} et~al.}{1998}]{1998MNRAS.299..554W}
{Whitworth} A.~P.,  {Boffin} H.~M.~J.,   {Francis} N.,  1998, \mn@doi [\mnras]
  {10.1046/j.1365-8711.1998.01813.x}, \href
  {http://adsabs.harvard.edu/abs/1998MNRAS.299..554W} {299, 554}

\bibitem[\protect\citeauthoryear{{Wiersma}, {Schaye}, {Dalla Vecchia}, {Booth},
  {Theuns}  \& {Aguirre}}{{Wiersma} et~al.}{2010}]{2010MNRAS.409..132W}
{Wiersma} R.~P.~C.,  {Schaye} J.,  {Dalla Vecchia} C.,  {Booth} C.~M.,
  {Theuns} T.,   {Aguirre} A.,  2010, \mn@doi [\mnras]
  {10.1111/j.1365-2966.2010.17299.x}, \href
  {http://adsabs.harvard.edu/abs/2010MNRAS.409..132W} {409, 132}

\bibitem[\protect\citeauthoryear{{Wise}, {Abel}, {Turk}, {Norman}  \&
  {Smith}}{{Wise} et~al.}{2012}]{2012MNRAS.427..311W}
{Wise} J.~H.,  {Abel} T.,  {Turk} M.~J.,  {Norman} M.~L.,   {Smith} B.~D.,
  2012, \mn@doi [\mnras] {10.1111/j.1365-2966.2012.21809.x}, \href
  {http://adsabs.harvard.edu/abs/2012MNRAS.427..311W} {427, 311}

\bibitem[\protect\citeauthoryear{{Wiseman}, {Schady}, {Bolmer}, {Kr{\"u}hler},
  {Yates}, {Greiner}  \& {Fynbo}}{{Wiseman} et~al.}{2016}]{2016arXiv160700288W}
{Wiseman} P.,  {Schady} P.,  {Bolmer} J.,  {Kr{\"u}hler} T.,  {Yates} R.~M.,
  {Greiner} J.,   {Fynbo} J.~P.~U.,  2016, preprint, \href
  {http://adsabs.harvard.edu/abs/2016arXiv160700288W} {} (\mn@eprint {arXiv}
  {1607.00288})

\bibitem[\protect\citeauthoryear{{Woosley} \& {Heger}}{{Woosley} \&
  {Heger}}{2007}]{2007PhR...442..269W}
{Woosley} S.~E.,  {Heger} A.,  2007, \mn@doi [\physrep]
  {10.1016/j.physrep.2007.02.009}, \href
  {http://adsabs.harvard.edu/abs/2007PhR...442..269W} {442, 269}

\bibitem[\protect\citeauthoryear{{Yajima}, {Nagamine}, {Thompson}  \&
  {Choi}}{{Yajima} et~al.}{2014}]{2014MNRAS.439.3073Y}
{Yajima} H.,  {Nagamine} K.,  {Thompson} R.,   {Choi} J.-H.,  2014, \mn@doi
  [\mnras] {10.1093/mnras/stu169}, \href
  {http://adsabs.harvard.edu/abs/2014MNRAS.439.3073Y} {439, 3073}

\bibitem[\protect\citeauthoryear{{Yajima}, {Shlosman}, {Romano-D{\'{\i}}az}  \&
  {Nagamine}}{{Yajima} et~al.}{2015}]{2015MNRAS.451..418Y}
{Yajima} H.,  {Shlosman} I.,  {Romano-D{\'{\i}}az} E.,   {Nagamine} K.,  2015,
  \mn@doi [\mnras] {10.1093/mnras/stv974}, \href
  {http://adsabs.harvard.edu/abs/2015MNRAS.451..418Y} {451, 418}

\bibitem[\protect\citeauthoryear{{Yamasawa}, {Habe}, {Kozasa}, {Nozawa},
  {Hirashita}, {Umeda}  \& {Nomoto}}{{Yamasawa}
  et~al.}{2011}]{2011ApJ...735...44Y}
{Yamasawa} D.,  {Habe} A.,  {Kozasa} T.,  {Nozawa} T.,  {Hirashita} H.,
  {Umeda} H.,   {Nomoto} K.,  2011, \mn@doi [\apj]
  {10.1088/0004-637X/735/1/44}, \href
  {http://adsabs.harvard.edu/abs/2011ApJ...735...44Y} {735, 44}

\bibitem[\protect\citeauthoryear{{Yan}, {Lazarian}  \& {Draine}}{{Yan}
  et~al.}{2004}]{2004ApJ...616..895Y}
{Yan} H.,  {Lazarian} A.,   {Draine} B.~T.,  2004, \mn@doi [\apj]
  {10.1086/425111}, \href {http://adsabs.harvard.edu/abs/2004ApJ...616..895Y}
  {616, 895}

\bibitem[\protect\citeauthoryear{{Zhou}, {Cao}  \& {Wu}}{{Zhou}
  et~al.}{2015}]{2015AJ....149....1Z}
{Zhou} Z.-M.,  {Cao} C.,   {Wu} H.,  2015, \mn@doi [\aj]
  {10.1088/0004-6256/149/1/1}, \href
  {http://adsabs.harvard.edu/abs/2015AJ....149....1Z} {149, 1}

\bibitem[\protect\citeauthoryear{{Zhukovska}}{{Zhukovska}}{2014}]{2014A&A...562A..76Z}
{Zhukovska} S.,  2014, \mn@doi [\aap] {10.1051/0004-6361/201322989}, \href
  {http://adsabs.harvard.edu/abs/2014A%26A...562A..76Z} {562, A76}

\bibitem[\protect\citeauthoryear{{Zhukovska}, {Gail}  \&
  {Trieloff}}{{Zhukovska} et~al.}{2008}]{2008A&A...479..453Z}
{Zhukovska} S.,  {Gail} H.-P.,   {Trieloff} M.,  2008, \mn@doi [\aap]
  {10.1051/0004-6361:20077789}, \href
  {http://adsabs.harvard.edu/abs/2008A%26A...479..453Z} {479, 453}

\bibitem[\protect\citeauthoryear{{Zhukovska}, {Dobbs}, {Jenkins}  \&
  {Klessen}}{{Zhukovska} et~al.}{2016}]{2016arXiv160804781Z}
{Zhukovska} S.,  {Dobbs} C.,  {Jenkins} E.~B.,   {Klessen} R.,  2016, preprint,
  \href {http://adsabs.harvard.edu/abs/2016arXiv160804781Z} {} (\mn@eprint
  {arXiv} {1608.04781})

\bibitem[\protect\citeauthoryear{{de Blok}, {Walter}, {Brinks}, {Trachternach},
  {Oh}  \& {Kennicutt}}{{de Blok} et~al.}{2008}]{2008AJ....136.2648D}
{de Blok} W.~J.~G.,  {Walter} F.,  {Brinks} E.,  {Trachternach} C.,  {Oh}
  S.-H.,   {Kennicutt} Jr. R.~C.,  2008, \mn@doi [\aj]
  {10.1088/0004-6256/136/6/2648}, \href
  {http://adsabs.harvard.edu/abs/2008AJ....136.2648D} {136, 2648}

\bibitem[\protect\citeauthoryear{{de Vaucouleurs} \& {Pence}}{{de Vaucouleurs}
  \& {Pence}}{1978}]{1978AJ.....83.1163D}
{de Vaucouleurs} G.,  {Pence} W.~D.,  1978, \mn@doi [\aj] {10.1086/112305},
  \href {http://adsabs.harvard.edu/abs/1978AJ.....83.1163D} {83, 1163}

\makeatother
\end{thebibliography}




\appendix

\section{Destruction of Newly Produced Dust}
\label{dustDestructionRate}

We consider SNe occurring in a newly formed stellar particle in the simulation.
The number of SNe is denoted as $N$, which is estimated in equation \eqref{floor} .
For simplicity,
we assume that each SN provides mass $\Delta M_{\rm d}^{(1)}$ of dust, 
which is evaluated as 
\begin{eqnarray} 
\Delta M_{\rm d}^{(1)}&=&
\dfrac{1}{N} \times f_{\rm in }
\mathcal{Y}^{\prime }_{\rm Z}\Delta M_{{\rm return}}\, .\label{YieldAppendix}
\end{eqnarray}
Since there is no spatially resolved information within each gas particle,
we assume that the dust produced by SNe is instantaneously distributed 
to the surrounding gas particles according to the kernel function.
The dust that was produced by the first SN in the gas particle 
would suffer from the shocks of subsequent $N-1$ SNe. 
In general, the dust produced by the $i$th SN in the particle 
experiences destruction by $N-i$ shocks.

As estimated in Section 2.2, 
each SN destroys a fraction $\varepsilon m_{\rm sw}\slash m_{\rm g}$ of dust 
within the gas particle.
Thus, if we define the surviving fraction as
$\eta_{\rm s} \equiv 1-\varepsilon_{\rm SN}{m_{\rm sw}}\slash{m_{\rm g}}$
($ 0 < \eta_{\rm s} < 1 $ in our simulation),
the net increase of dust after destruction is
\begin{eqnarray} 
\Delta M^{\prime }_{\rm d}&=&\Delta M^{(1)}_{\rm d}
\displaystyle\sum_{k=0}^{N-1}\eta_{\rm s}^{k} 
= \begin{cases}
\Delta M_{\rm d}^{(1)}\dfrac{1-\eta_{\rm s}^{N}}{1-\eta_{\rm s}}~({\rm for~}N \ge 2)~,\\
\Delta M_{\rm d}^{(1)}~({\rm for~}N = 1)~.
\end{cases}
\label{survivalRate}
\end{eqnarray}
By using equation \eqref{YieldAppendix}, we obtain $\Delta M^{\prime }_{\rm d}$ as
\begin{eqnarray} 
\Delta M^{\prime }_{\rm d}&=&
\begin{cases}
\dfrac{1}{N}\times f_{\rm in}
\mathcal{Y}^{\prime }_{\rm Z}\Delta M_{{\rm return}}
\dfrac{1-\eta_{\rm s}^{N}}{1-\eta_{\rm s}}~({\rm for~}N \ge 2),\\
f_{\rm in}\mathcal{Y}^{\prime }_{\rm Z}\Delta M_{{\rm return}}~({\rm for~}N = 1).
\end{cases}
\label{survivalRate2}
\end{eqnarray}

On the other hand, $\Delta M_{\rm d}^{\prime }$ can also be written 
with the destroyed fraction $\delta $ as 
\begin{eqnarray} 
\Delta M_{\rm d}^{\prime }&=&f_{\rm in}\mathcal{Y}^{\prime }_{\rm Z}\Delta M_{{\rm return}}( 1 - \delta )\, . \label{YieldAppendix2}
\end{eqnarray}
By combining equation \eqref{survivalRate2} with equation \eqref{YieldAppendix2}, we obtain
\begin{eqnarray} 
(1-\delta)&=&
\begin{cases}
\dfrac{1}{N}\dfrac{1-\eta^{N}_{\rm s}}{1-\eta_{\rm s}}~({\rm for~}N \ge 2 )\,,\\
1~({\rm for~}N=1)\, .
\end{cases}
\end{eqnarray}

\section{Derivation of the time-scale of collisional process}\label{collisional}
For grain--grain collisional processes such as coagulation and shattering, 
the time-scales are estimated based on the collision time-scale $\tau_\mathrm{coll}$.
This time-scale can be estimated by
\begin{eqnarray}
\tau_{\rm coll}=\dfrac{1}{\sigma vn_{\rm d}}~,
\end{eqnarray}
where $v$, $\sigma$ and $n_{\rm d}$ are the typical velocity
dispersion, the cross-section and the number density of the dust
grains, respectively. The cross-section can be estimated as
\begin{eqnarray} 
\sigma \simeq \pi a^{2}~,
\end{eqnarray}
where $a$ is the grain radius 
(we assume spherical grains).
For convenience, we express
$n_\mathrm{d}$ with the dust-to-gas ratio,
$\mathcal{D}$,
\begin{eqnarray}
n_\mathrm{d}=\frac{\mathcal{D}\mu m_\mathrm{H}n_\mathrm{H}}{\frac{4}{3}\pi a^3s},
\end{eqnarray}
where $\mu =1.4$ is the mean weight of gas particles per hydrogen,
$m_\mathrm{H}$ is the hydrogen mass (i.e.\ $\mu m_\mathrm{H}$ is
the gas mass per hydrogen nucleus), and $s$ is the material
density of the dust. Using the dust-to-gas ratio,
the collision time-scale is expressed as
\begin{eqnarray}
\tau_\mathrm{coll} &=&
\frac{\frac{4}{3}as}{v\mathcal{D}\mu m_\mathrm{H}n_\mathrm{H}}
\nonumber \\
&=&5.41 \times 10^{7}
\left(\frac{\mathcal{D}}{0.01}\right)^{-1}
\left(\frac{n_\mathrm{H}}{1~\mathrm{cm}^{-3}}\right)^{-1}
\left(\frac{a}{0.1~\mu\mathrm{m}}\right)\notag \\
& &\times \left(\frac{s}{3~\mathrm{g}~\mathrm{cm}^{-3}}\right)
\left(\frac{v}{10~\mathrm{km}~\mathrm{s}^{-1}}\right)^{-1}
\mathrm{[yr]}~.
\label{eq:tau_coll}
\end{eqnarray}

\bsp	
\label{lastpage}
\end{document}